\documentclass[aps,prb,reprint,superscriptaddress]{revtex4-1}
\usepackage{blindtext}
\usepackage{color}
\usepackage{dcolumn}
\usepackage{amsmath}
\usepackage{amssymb}
\usepackage{mathtools}
\usepackage{graphicx}
\usepackage{latexsym}
\usepackage{amsfonts}
\usepackage{amssymb}
\usepackage{comment}
\usepackage{natbib}
\usepackage{graphicx}
\usepackage{soul}

\definecolor{crimson}{RGB}{255,102,255}

\begin{document}

\title{Photon-assisted stochastic resonance in nanojunctions}
\author{Michael Ridley}
\affiliation{Faculty of Engineering and Institute of Nanotechnology and Advanced Materials,
Bar Ilan University, Ramat Gan 52900, Israel}
\email{mikeridleyphysics@gmail.com}

\author{Leo Bellassai}
\affiliation{Department of Physics, Nanoscience Center, P.O. Box 35, 40014 University of Jyv{\"a}skyl{\"a}, Finland}

\author{Michael Moskalets}
\affiliation{Institute for Cross-Disciplinary Physics and Complex Systems IFISC (UIB-CSIC), 07122 Palma de Mallorca, Spain}
\affiliation{Department~of~Metal~and~Semiconductor~Physics,~NTU~``Kharkiv Polytechnic Institute'', 61002~Kharkiv,~Ukraine \looseness=-1}

\author{Lev Kantorovich}
\affiliation{Theory and Simulation of Condensed Matter (TSCM), King's College London, Strand, London WC2R 2LS, United Kingdom \looseness=-1}

\author{Riku Tuovinen}
\affiliation{Department of Physics, Nanoscience Center, P.O. Box 35, 40014 University of Jyv{\"a}skyl{\"a}, Finland}

\begin{abstract}

%We study stochastic resonance in molecular structures driven by a periodically-varying external driving field. This is done using the time-dependent Landauer-B{\"u}ttiker formalism, which follows from exact analytical solutions to the Kadanoff-Baym equations describing a nanojunction subject to an arbitrary time-dependent bias. We apply this formalism to a double quantum dot nanojunction, and compare the effects of the temperature with the fluctuating bias in the non-driven case. We then consider the combined effect of AC-driving and white noise fluctuations on the rectified current through the molecule, and find a novel stochastic resonance effect in the signal. The study is then extended to include the color noise, so that the combined effect of the color noise correlation time and driving frequency on stochastic resonance is calculated. We thereby demonstrate that photon-assisted transport can be optimized by a suitably tuned environment.

We study stochastic resonance in molecular junctions driven by a periodically-varying external field. This is done using the time-dependent Landauer-B{\"u}ttiker formalism, which follows from exact analytical solutions to the Kadanoff-Baym equations describing the molecular junction  subject to an arbitrary time-dependent bias. We focus on a double quantum dot nanojunction and compare the effects of the temperature with the fluctuating bias in the statically-driven case. We then consider the combined effect of AC-driving and white noise fluctuations on the rectified current through the nanojunction, and find a stochastic resonance effect, where at certain driving conditions the bias fluctuations enhance the current signal. The study is then extended to include the color noise in the applied bias, so that the combined effect of the color noise correlation time and driving frequency on stochastic resonance is investigated. We thereby demonstrate that photon-assisted transport can be optimized by a suitably tuned environment.
\end{abstract}

\maketitle

\section{Introduction}
In most signal processing systems, environmental noise is viewed as a hindrance to the clean transfer of information. However, this preconception was turned on its head by the discovery of stochastic resonance (SR), referring to the counterintuitive property of signal enhancement in the presence of a suitable level of external noise \cite{gammaitoni1998stochastic}. Initially coined to describe the periodic recurrence of ice ages on Earth \cite{benzi1981mechanism}, SR has been studied in a diverse range of dynamical systems, with applications in electrical engineering \cite{luchinsky1999stochastic,harmer2002review,mikhaylov2021stochastic}, 
chemistry \cite{yilmaz2013stochastic,yamakou2020optimal}, paleoclimatology \cite{alley2001stochastic,ganopolski2002abrupt} and biology \cite{douglass1993noise,hanggi2002stochastic,mcdonnell2009stochastic}. The phenomenon has also been studied in various quantum model systems \cite{lofstedt1994quantum,grifoni1996coherent,goychuk1999quantum}, and in structures including the spin-boson model \cite{huelga2007stochastic}, scanning tunneling microscopes \cite{hanze2021quantum} and Rydberg atoms \cite{li2024collective}. In addition, recent experimental work \cite{wagner2019quantum} indicates that this phenomenon can be observed in driven nanoscale systems, where the fluctuations of the environment can have a profound effect on the operation of electrical devices \cite{german1994resonant,dykman1995statistical,astumian1998overview,jiang2010control,hartmann2010stochastic,soni2010probing,bargueno2011quantum,hayashi2012single,popescu2012time,hirano2013stochastic,brunner2014random,pfeffer2015logical,ridley_fluctuating-bias_2016,gurvitz2016temporal,fujii2017single,entin2017heat,yoshida2017stochastic,kosov2018telegraph,aharony2019telegraph,hussein2020spectral,markina2020detection,roldan2024stochastic}.  

Single molecule junctions are increasingly used as an analytical technique for the investigation of strongly nonequilibrium properties of circuit components \cite{tao2006electron,li2023molecular}. Recent experimental work has focused on the dynamical
properties of nanojunctions in the GHz-THz frequency range \cite{yoshioka2016real,garg2020attosecond,viti2020thermoelectric,qiu2021photodetectors,leitenstorfer20232023} and light-driven electron transport \cite{li2024s,sun2024development}. However, although nanoscale junctions are particularly susceptible to fluctuations in their environment, to date there has been no fully quantum treatment of SR in periodically-driven nanodevices, where effects such as photon-assisted tunneling (PAT) come into play \cite{moskalets2002floquet,arrachea2006relation,iurov2011anomalous,moskalets2011scattering,ridley_time-dependent_2017}. 

\begin{figure}[t]
%\centering
\includegraphics[width=0.4\textwidth]{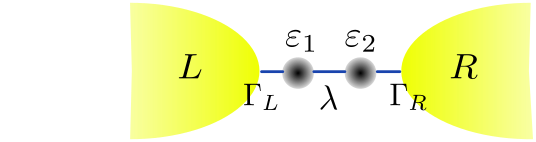}
\caption{Schematic illustration of a DQD nanojunction with dot energy levels $\varepsilon_{1}$, $\varepsilon_{2}$ coupled to the left ($L$) and right ($R$) leads. The level broadening due to the two leads is $\Gamma_{L}$, $\Gamma_{R}$ respectively and $\lambda$ is the inter-dot coupling.} 
\label{DQD}
\end{figure}

An important species of nanodevice is the double quantum dot (DQD) nanojunction, illustrated schematically in Fig.~\ref{DQD}. Quantum dot and DQD semiconductor structures find many uses in quantum information processing, as they can be used to create charge \cite{petta2004manipulation,gorman2005charge,nielsen2013six, scarlino2022situ} and spin \cite{petta2005coherent,hanson_spins_2007,fernandez2022quantum,burkard2023semiconductor} qubits with long coherence times, making them ideal hardware for quantum computing implementations.

Nanoelectronic systems like the DQD nanojunction are well-described by the nonequilibrium Green's function (NEGF) technique, an established method for describing quantum transport~\cite{ridley2022many}. In particular, the NEGF method provides access to conductance properties~\cite{wang1999current}, time-dependent currents~\cite{you2000landauer,tuovinen_time-dependent_2013}, finite temperature effects~\cite{baldea2024can} and shot noise~\cite{ridley_partition-free_2017}, as well as higher moments of the full counting statistics~\cite{tang_full-counting_2014} in response to a 
perturbation driving the system out of equilibrium.
%quench at the switch-on time $t_{0}$. 

In the general case, the NEGF approach can simultaneously deal with strong external fields, many-particle interactions, and transient effects for various quantum-transport setups~\cite{ridley2022many,tang2024modulating}. While the drawback of this approach resides in the computational effort for solving the dynamical equations of motion, an accurate but computationally tractable scheme can be formulated in terms of the generalized Kadanoff-Baym ansatz~\cite{Lipavsky1986,Latini2014,Tuovinen2021}. Recently, it has been established that this reconstruction results in a time-linear computational scheme for electron-electron~\cite{Schluenzen2020,Joost2020}, electron-boson~\cite{Karlsson2021,tuovinen2024electroluminescence}, and embedding effects~\cite{tuovinen_time-linear_2023, pavlyukh_cheers_2024, cosco_interacting_2024}. However, in many experimentally and technologically relevant settings involving nanojunctions, a noninteracting approach can be sufficient~\cite{Dutta2020, Ojajarvi2022, Ryu2022}.
%, as shall be discussed next.

A recent development of an NEGF-based approach, the so-called time-dependent Landauer-B{\"u}ttiker (TD-LB) formalism~\cite{stefanucci_nonequilibrium_2013,tuovinen_time-dependent_2013,tuovinen_time-dependent_2014,ridley_current_2015,ridley_calculation_2016,ridley_formal_2018,ridley2022many}, has been found to yield exact expressions for the time-dependent currents and noise in nanojunctions of arbitrary size, temperature and any time-dependent driving fields. In addition, this formalism incorporates the \emph{partition-free} approach~\cite{stefanucci2004time,ridley_formal_2018}, where the lead-molecule coupling is present in the Hamiltonian at equilibrium, prior to the switch-on of a bias.
%In \emph{partitioned} approaches, such as that taken in Ref. \cite{jauho_time-dependent_1994}, the lead-molecule coupling and the external bias are simultaneously added to the Hamiltonian at $t_{0}$.
The partition-free approach thus gives a more experimentally realizable description of the transient regime following a voltage quench in the nanojunction. 
The partition-free TD-LB methodology has recently been applied to superconducting nanowires~\cite{tuovinen_time-dependent_2016,tuovinen_distinguishing_2019}, graphene nanoribbon structures~\cite{ridley_time-dependent_2017,ridley_electron_2019} and the local radiation profiles of molecular junctions~\cite{ridley_quantum_2021}. In addition, the TD-LB was used to study the problem of a driving field with a stochastic component in Ref.~\onlinecite{ridley_fluctuating-bias_2016}, in which a quantum version of the classical Nyquist theorem relating conductance to the field fluctuations and temperature was derived.

In this work, we apply the TD-LB approach to the study of SR in a periodically driven nanojunction. The paper is structured as follows. In Section~\ref{sec:model}, we outline the quantum transport setup for a nanojunction driven by biharmonic and Gaussian stochastic bias contributions. Section~\ref{sec:TDLB} contains the main equations of the TD-LB formalism for an arbitrary time-dependent bias, supported by the underlying NEGF theory. In Section~\ref{sec:noisybias}, the resulting current formulas, following an averaging procedure over the stochastic bias, are expressed in a computationally tractable form in which all time and frequency integrals are calculated analytically yielding the current expression in a closed form via sums of special functions. The resulting expression is then averaged over a period of the driving field to give the net pump current across the nanojunction. The corresponding adiabatic limit is considered explicitly for the case of a quantum dot junction subject to a white noise bias. In Section~\ref{sec:numerics}, numerical results are presented for the DQD system shown in Fig.~\ref{DQD}. The main findings are: the fluctuation-dependence of current-voltage characteristics for the static bias case; the activation of photon-assisted SR; and the effects of particle-hole symmetry-breaking and color noise on the SR phenomenon. Finally, in Section~\ref{sec:concl} we conclude and provide an outlook.

\section{Model and method}

%\textcolor{red}{Notation: (remove this)
%\begin{align}
%\phi_\alpha & \quad \text{phase difference in the biharmonic voltage profile} \nonumber \\
%\psi_\alpha & \quad \text{bias-voltage phase factor (time integral) } \nonumber \\
%\varphi_i^{L/R} & \quad \text{left-right eigenvectors of the effective Hamiltonian} \nonumber \\
%\varepsilon_i & \quad \text{on-site potential} \nonumber \\
%\bar{\varepsilon}_i & \quad \text{eigenvalue of the effective Hamiltonian} \nonumber
%\end{align}
%Suggestion to have a different notation for on-site potential, since we already use epsilon ($\bar{\varepsilon}$) for the eigenvalues, and it appears in numerous formula and is more difficult to change. Alternatively, write a note after Eq.~\eqref{eq:h_eff_LR} to comment on the difference.
%}

\subsection{Quantum transport setup}\label{sec:model}
For a typical set-up of a quantum transport problem shown in Fig.~\ref{DQD}, we use the following Hamiltonian:
\begin{align}\label{eq:Hamiltonian}
    \hat{H}\left(z\right)=\underset{k\alpha\sigma}{\sum}\varepsilon_{k\alpha}\left(z\right)\hat{d}_{k\alpha\sigma}^{\dagger}\hat{d}_{k\alpha\sigma}+\underset{mn\sigma}{\sum}h_{mn}\left(z\right)\hat{d}_{m\sigma}^{\dagger}\hat{d}_{n\sigma}\nonumber\\
    +\underset{m,k\alpha\sigma}{\sum}\left[T_{mk\alpha}\left(z\right)\hat{d}_{m\sigma}^{\dagger}\hat{d}_{k\alpha\sigma}+T_{k\alpha m}\left(z\right)\hat{d}_{k\alpha\sigma}^{\dagger}\hat{d}_{m\sigma}\right],
\end{align}
where $z$ is a time parameter on a contour, which is addressed in more detail in Sec.~\ref{sec:TDLB}. In Eq.~\eqref{eq:Hamiltonian},
the first term corresponds to the sum of the Hamiltonians of the reservoirs/leads, where $\alpha$ labels the leads, and $k$ labels the $k$-th eigenstate of a lead. The second term corresponds to the Hamiltonian of the central region $C$ (the DQD in our case) and hence refers to electron hopping events within this region with indices $n$ and $m$ labeling localised orbitals there. Finally, the third term describes the coupling of the leads and the central system with the corresponding matrix elements $T_{mk\alpha}$, and $\sigma$ denotes the spin degree of freedom of the electrons. Correspondingly, $\hat{d}_{k\alpha\sigma}$, $\hat{d}_{m\sigma}$ and $\hat{d}_{k\alpha\sigma}^{\dagger}$, $\hat{d}_{m\sigma}^{\dagger}$ are destruction and creation operators of the leads and the central system. Note that there is no direct interaction between the leads.

The system is driven out of equilibrium by the switch-on of a spatially uniform bias $V_{\alpha}\left(t\right)$ in each reservoir, modifying their energy dispersion $\varepsilon_{k\alpha}\to \varepsilon_{k\alpha} + V_\alpha(t)$. We assume this bias to be broken up into a constant shift $V_{\alpha}$, a deterministic time-dependent driving 
\begin{align}\label{eq:driving}
V_{\alpha}^d\left(t\right) &=A_{\alpha}^{\left(1\right)}\cos\left(p_{1}\Omega_{\alpha}\left(t-t_{0}\right)+\phi_{\alpha}\right) \nonumber\\
&+A_{\alpha}^{\left(2\right)}\cos\left(p_{2}\Omega_{\alpha}\left(t-t_{0}\right)\right)
\end{align}
chosen in the biharmonic form, 
%the biharmonic external field $V_{\alpha}^{D}\left(t\right)$ 
and a stochastic time-dependent field $V_{\alpha}^{s}\left(t\right)$ leading to the full time-dependent bias being
\begin{align}\label{eq:drivingprotocol}
    V_{\alpha}\left(t\right)=V_{\alpha}+V_{\alpha}^{d}\left(t\right)+V_{\alpha}^{s}\left(t\right).
\end{align}
%where we choose the biharmonic external field 
The bias voltage protocol $V_\alpha(t)$ could be realized by an external electromagnetic field, and from now on it will be generally referred to as ``photon-assisted'' driving.
We further assume that $V_{\alpha}^{s}\left(t\right)$ is a zero-mean, stationary, Gaussian stochastic process, uniform across the leads so that we can drop the index $\alpha$. The Gaussian nature of the noise  means that odd-ordered statistical moments vanish, while  even-ordered moments can be decomposed into a sum of products of pair correlation functions: 
\begin{align}\label{eq:1moment}
    \overline{V^{s}\left(t\right)}=0
\end{align}
\begin{align}
\overline{V^{s}\left(t_{1}\right)V^{s}\left(t_{2}\right)}=\overline{V^{s}\left(t_{1}-t_{2}\right)V^{s}\left(0\right)}=C\left(t_{1}-t_{2}\right)
\end{align}
\begin{align}
    \overline{V^{s}\left(t_{1}\right)\ldots V^{s}\left(t_{2n}\right)}=\underset{p}{\sum}C\left(t_{p_{1}}-t_{p_{2}}\right)\ldots C\left(t_{p_{2n-1}}-t_{p_{2n}}\right)
\end{align}
\begin{align}
    \overline{V^{s}\left(t_{1}\right)\ldots V^{s}\left(t_{2n+1}\right)}=0,
\end{align} 
where we introduced a bias correlation function $C$, ${\sum_p}$ denotes a summation over all permutations of pairs of the time variables, and the bar denotes the average taken with respect to the stochastic distribution of the bias. In analogy with classical approaches \cite{pottier2009nonequilibrium}, we choose a special form of the correlation function which involves a parameter $\omega_{c}=1/\tau_{c}$, defining a finite correlation time $\tau_{c}$, over which the bias is statistically correlated:
\begin{align}\label{eq:correlation}
    C\left(\tau_{1}-\tau_{2}\right)=D\omega_{c}e^{-\omega_{c}\left|\tau_{1}-\tau_{2}\right|},
\end{align}
where $D$ is a parameter measuring the overall magnitude of bias fluctuations, such that ${\int}_{-\infty}^\infty d\tau C\left(\tau\right)=2D$. One intuitively expects an increase in $D$ to increase the dissipation and dephasing of molecular eigenmodes, as was discussed previously in Ref.~\onlinecite{ridley_fluctuating-bias_2016}.

\subsection{Time-dependent Landauer-B{\"u}ttiker formalism}\label{sec:TDLB}
%We now provide a brief overview of the TD-LB method. 
Based on the solution to the Kadanoff-Baym equations~\cite{kadanoff_quantum_1962}, the NEGF technique provides formally exact access to the one-particle Green's function
\begin{align}\label{eq:GF1}
G_{1}\left(1;2\right)\equiv\frac{\textrm{Tr}\left[\hat{T}_{\gamma}\left[e^{-i\int_{\gamma}d\bar{z}\hat{H}\left(\bar{z}\right)}\hat{\psi}\left(1\right)\hat{\psi}^{\dagger}\left(2\right)\right]\right]}{\textrm{Tr}\left[\hat{T}_{\gamma}\left[e^{-i\int_{\gamma}d\bar{z}\hat{H}\left(\bar{z}\right)}\right]\right]},
\end{align}
where 
%$\beta$ denotes the inverse temperature, 
the $\hat{\psi}$ are fermionic field operators and $\hat{H}\left(\bar{z}\right)$ denotes the Hamiltonian defined with respect to time variables $z$ on the Konstantinov-Perel' time contour $\gamma$ %consisting of an `upper' time branch $C^{f}$ running from $t_{0}$ to $t$, a `lower' time branch %$C^{b}$ running backwards from $t$ to $t_{0}$, with a vertical Matsubara branch $C^{M}$ running from %$t_{0}$ to $t_{0}-i\beta$ (\mm{may be it is better to say here what is \(\\beta)\), not at the %beginning of the paragraph?}) %\cite{konstantinov_graphical_1960,keldysh_diagram_1964,stefanucci_nonequilibrium_2013}. 
%Here we use the full Konstantinov-Perel's time contour $\gamma$ 
broken into $\gamma_{-}$ (running from $t_0$ to $t$), $\gamma_{+}$ (running backwards from $t$ to $t_0$) and $\gamma_{M}$ (running from $t_0$ to $t_0-i\beta$, where $\beta$ is the inverse temperature)~\cite{konstantinov_graphical_1960, keldysh_diagram_1964, stefanucci_nonequilibrium_2013}.
The variables $i=1,2$ correspond to spacetime locations $(x_{i},z_{i})$ with $z_{i}\in\gamma$ and $x_i$ being the spatial variable including the spin.

We collect the elements of the Hamiltonian Eq.~\eqref{eq:Hamiltonian} into the block matrix $\mathbf{h}\left(z\right)$. A full description of the dynamics subsequent to the bias switch-on requires a specification of $\mathbf{h}\left(z\right)$ at every contour time $z$:
\begin{align}
\left[\mathbf{h}_{\alpha\alpha}\left(z\right)\right]_{kk'} & = \begin{cases} \left(\varepsilon_{k\alpha}+V_{\alpha}\left(t\right)\right)\delta_{kk'}, & z\equiv t\in \gamma_{-}\oplus \gamma_{+}\\
\left(\varepsilon_{k\alpha}-\mu\right)\delta_{kk'}, & z\in \gamma_{M}\end{cases}\\
\left[\mathbf{h}_{C\alpha}\left(z\right)\right]_{mk}&=T_{mk\alpha}, \quad z\in\gamma \\
\left[\mathbf{h}_{CC}\left(z\right)\right]_{mn}
&=\begin{cases}
h_{mn}, & z\in \gamma_{-}\oplus \gamma_{+}\\
h_{mn}-\mu\delta_{mn}, & z\in \gamma_{M},
\end{cases}
\end{align}
where $\mu$ is the chemical potential and $C$ refers to the central region.

The Green's function in Eq.~\eqref{eq:GF1} is projected onto the central (molecular) region to obtain the matrix-valued function $\mathbf{G}_{CC}$, which satisfies the Kadanoff-Baym integro-differential equations of motion \cite{stefanucci_nonequilibrium_2013}
\begin{align}
&\left[i\frac{d}{dz_{1}}-\mathbf{h}_{CC}\left(z_{1}\right)\right]\mathbf{G}_{CC}\left(z_{1},z_{2}\right)\nonumber \\
&=\mathbf{1}_{CC}\delta\left(z_{1},z_{2}\right)
+\int_{\gamma}d\bar{z}\,\mathbf{\Sigma}_{em}\left(z_{1},\bar{z}\right)\mathbf{G}_{CC}\left(\bar{z},z_{2}\right),\label{eq:kbe1}\\
&\mathbf{G}_{CC}\left(z_{1},z_{2}\right)\left[-i\frac{d}{dz_{2}}-\mathbf{h}_{CC}\left(z_{2}\right)\right]\nonumber \\
&=\mathbf{1}_{CC}\delta\left(z_{1},z_{2}\right)
+\int_{\gamma}d\bar{z}\
\mathbf{G}_{CC}\left(z_{1},\bar{z}\right)\mathbf{\Sigma}_{em}\left(\bar{z},z_{2}\right)\label{eq:kbe2},
\end{align}
with the integral kernel given by the embedding self-energy
\begin{align}
\left[\mathbf{\Sigma}_{em}\left(z_{1},z_{2}\right)\right]_{mn}&=\underset{k \alpha}{\sum}T_{mk\alpha}\left[\mathbf{g}_{\alpha\alpha}\left(z_{1},z_{2}\right)\right]_{kk}T_{k\alpha n}\nonumber\\
&\equiv \underset{ \alpha}{\sum}\left[\mathbf{\Sigma}_{em,\alpha}\left(z_{1},z_{2}\right)\right]_{mn},
\end{align}
where $\mathbf{g}_{\alpha\alpha}$ is the Green's function of the decoupled lead $\alpha$. Eqs. \eqref{eq:kbe1} and \eqref{eq:kbe2} are then projected onto equations for different components of the Green's function (corresponding to different combinations of pairs of contour branch times) using the Langreth rules\cite{langreth1976linear}. 

We now assume that the leads satisfy the wide-band limit approximation (WBLA), i.e. we neglect the energy dependence of the lead-molecule coupling. This assumption enables us to write down all components of the effective embedding self-energy in terms of the {\em energy-independent} level-width matrix
\begin{align}
    \Gamma_{\alpha,mn}=2\pi\sum_{k}T_{m k\alpha}T_{k\alpha n}\delta\left(\varepsilon_{\alpha}^{F}-\varepsilon_{k\alpha}\right),
\end{align}
where $\varepsilon_{\alpha}^{F}$ is the equilibrium Fermi energy of lead $\alpha$. 
Within the WBLA, Eqs.~\eqref{eq:kbe1} and~\eqref{eq:kbe2} are linearized in terms of the effective Hamiltonian of the central region,
$\mathbf{h}_{CC}^{\text{eff}}\equiv\mathbf{h}_{CC}-\frac{i}{2}{\sum_\alpha} \boldsymbol{\Gamma}_{\alpha}$.
All components of the Green's function can be calculated exactly in the two-time plane \cite{ridley_current_2015,ridley_fluctuating-bias_2016}. These Green's function components and the corresponding embedding self-energy components are listed in Appendix \ref{sec:Appendix_A}.

\begin{widetext}
The quantum statistical expectation value of the current operator is given by
\begin{equation}
    \label{eq:current_1}
    I_{\alpha}\left(t\right)=4q\textrm{Re}\textrm{Tr}_{C}\left[\mathbf{h}_{C\alpha}\mathbf{G}_{\alpha C}^{<}\left(t,t\right)\right]\\
    =-4\textrm{Re}\textrm{Tr}_{C}\left[\left(\mathbf{\Sigma}_{em,\alpha}^{<}\cdot\mathbf{G}_{CC}^{a}+\mathbf{\Sigma}_{em,\alpha}^{r}\cdot\mathbf{G}_{CC}^{<}+\mathbf{\Sigma}_{em,\alpha}^{\rceil}\star\mathbf{G}_{CC}^{\lceil}\right)\left(t,t\right)\right],
\end{equation}
where $\cdot$ and $\star$ correspond to real and imaginary time convolutions \cite{stefanucci_nonequilibrium_2013}.
Setting the electronic charge $q=-1$ and using the derived formulae for the components of $\mathbf{G}_{CC}$ in the WBLA, Eq.~\eqref{eq:current_1} can be shown to take the form~\cite{ridley_partition-free_2017}:
\begin{align}\label{eq:Current_alpha}
I_{\alpha}\left(t\right) & = \frac{1}{\pi}\int d\omega f\left(\omega-\mu\right)\,\mbox{Tr}_{C}\left[2\mbox{Re}\left[i\boldsymbol{\Gamma}_{\alpha}e^{i\omega\left(t-t_{0}\right)}e^{i\psi_{\alpha}\left(t,t_{0}\right)}\mathbf{S}_{\alpha}\left(t,t_{0};\omega\right)\right]-\boldsymbol{\Gamma}_{\alpha}\underset{\beta}{\sum}\mathbf{S}_{\beta}\left(t,t_{0};\omega\right)\boldsymbol{\Gamma}_{\beta}\mathbf{S}_{\beta}^{\dagger}\left(t,t_{0};\omega\right)\right] .
\end{align}
where $f\left(x\right)=\left(e^{\beta x}+1\right)^{-1}$ is the Fermi-Dirac distribution, and we have introduced the $\mathbf{S}_{\alpha}$ matrix
\begin{align}\label{eq:S_alpha}
& \mathbf{S}_{\alpha}\left(t,t_{0};\omega\right)\equiv e^{-i\mathbf{h}_{CC}^{\text{eff}}\left(t-t_{0}\right)}\left[\mathbf{G}_{CC}^{r}\left(\omega\right) -i\int_{t_{0}}^{t}d\bar{t}e^{-i\left(\omega\mathbf{1}-\mathbf{h}_{CC}^{\text{eff}}\right)\left(\bar{t}-t_{0}\right)}e^{-i\psi_{\alpha}\left(\bar{t},t_{0}\right)}\right]
\end{align}
defined in terms of the retarded Green's function $\mathbf{G}_{CC}^{r}\left(\omega\right)=\left(\omega\mathbf{I}-\mathbf{h}_{CC}^{\text{eff}}\right)^{-1}$. In Eq.~\eqref{eq:S_alpha}, we also introduced the bias-voltage phase factor
\begin{align}
\psi_{\alpha}\left(t_{1},t_{2}\right)\equiv{\int}_{t_2}^{t_1}d\tau\,V_{\alpha}\left(\tau\right).
\end{align}
It is essential that the bias enters only into these phase factors; hence, all the time information of the bias, including its stochastic part, is contained exclusively in exponential functions $\exp\left(\pm i\psi_{\alpha}(t,t_0) \right)$ that appear linearly everywhere in the current expression. 

Eq.~\eqref{eq:Current_alpha} is a generalization of the well-known \cite{landauer1957spatial,buttiker1986four} Landauer-B{\"u}ttiker formula for the current to include transient effects due to the partition-free quench, and due to an arbitrary time-dependent bias. We also note that the currents in the leads obey the following time-dependent continuity equation \cite{ridley_calculation_2016}
\begin{equation}\label{eq:continuity}
    \frac{d\mathbf{N}_{C}\left(t\right)}{dt}=\underset{\alpha}{\sum}I_{\alpha}\left(t\right),
\end{equation}
where $\mathbf{N}_{C}\left(t\right)=-2i\textrm{Tr}_{C}\left[\mathbf{G}_{CC}^{<}\left(t,t\right)\right]$ is the particle number in the central region and the lesser Green's function $\mathbf{G}_{CC}^{<}\left(t,t\right)$ is given in Eq.~\eqref{eq:GF_lessgtr}.

\subsection{Noise-averaging and pump current}\label{sec:noisybias}

For the practical purposes of calculating the current in Eq.~\eqref{eq:Current_alpha}, it is advantageous to analytically calculate all frequency and time integrals to express the current in terms of special functions. In addition, one has to sample the current over the Gaussian fluctuations of the bias. 
%In the model under consideration, the time-dependent voltage in the leads is contained in exponents of phase factors of the form 
%\begin{align}
%\psi_{\alpha}\left(t_{1},t_{2}\right)\equiv{\int}_{t_2}^{t_1}d\tau\,V_{\alpha}\left(\tau\right),
%\end{align}
%which are decomposed into a product of terms with stochastic and biharmonic origin.
The bias-voltage phase factors $\psi_\alpha$ appearing in Eq.~\eqref{eq:Current_alpha} may be decomposed into a product of terms with stochastic and biharmonic origin.
Because the bias voltage appears only in the exponent, its averaging over the stochastic part of the bias can be done analytically using the fact that the average of the stochastic phase factor over a Gaussian noise is given by
\begin{equation}\label{eq:phase_factor_correlation}
\overline{\exp\left(\pm i\int_{t_{1}}^{t_{2}}d\tau V^{s}\left(\tau\right)\right)} =\exp\left[-\frac{1}{2}\int_{t_{1}}^{t_{2}}d\tau_{1}\int_{t_{1}}^{t_{2}}d\tau_{2}C\left(\tau_{1}-\tau_{2}\right)\right],
\end{equation}
so that we can insert here the color noise correlation function from Eq. \eqref{eq:correlation} and expand this with respect to the stochastic bias in powers of $D\tau_{c}$, as was done in Ref.~\onlinecite{ridley_fluctuating-bias_2016}. We note that dealing with the stochastic contribution to the bias in this way removes the need to sample over multiple stochastic trajectories and therefore drastically reduces the computational cost associated with the noisy bias. In addition, we expand the exponentiated  harmonic terms via Bessel functions of the first kind $J_{n}\left(x\right)$ \cite{ridley_fluctuating-bias_2016,ridley_time-dependent_2017}, so that phase factors in Eq. \eqref{eq:Current_alpha} are replaced by:
\begin{align}\label{eq:phase_factor}
\overline{e^{i\psi_{\alpha}\left(t_{1},t_{2}\right)}}&=e^{iV_{\alpha}\left(t_{1}-t_{2}\right)}\underset{n=0}{\overset{\infty}{\sum}}\,\underset{m=0}{\overset{n}{\sum}}\,\underset{rr'ss'}{\sum}\left(\frac{D}{\omega_{c}}\right)^{n}\frac{\left(-1\right)^{m}}{\left(n-m\right)!m!}J_{rr'ss'}^{(\alpha)}
\nonumber \\
&\times e^{-\left(D+m\omega_{c}\right)\left|t_{2}-t_{1}\right|}e^{i\left(r-r'\right)\phi_{\alpha}}e^{i\Omega_{\alpha}\left(p_{1}r+p_{2}s\right)\left(t_{1}-t_{0}\right)}e^{-i\Omega_{\alpha}\left(p_{1}r'+p_{2}s'\right)\left(t_{2}-t_{0}\right)},
\end{align}
where
\begin{align}
J_{rr'ss'}^{(\alpha)}=J_{r}\left(\frac{A_{\alpha}^{\left(1\right)}}{p_{1}\Omega_{\alpha}}\right)J_{r'}\left(\frac{A_{\alpha}^{\left(1\right)}}{p_{1}\Omega_{\alpha}}\right)J_{s}\left(\frac{A_{\alpha}^{\left(2\right)}}{p_{2}\Omega_{\alpha}}\right)J_{s'}\left(\frac{A_{\alpha}^{\left(2\right)}}{p_{2}\Omega_{\alpha}}\right)\,.\label{eq:J}
\end{align}

These may be inserted into the expanded version of Eq.~\eqref{eq:Current_alpha}
\begin{align}\label{eq:I_expand}
\overline{I_{\alpha}\left(t\right)}&=\frac{1}{\pi}\int d\omega f\left(\omega-\mu\right)\textrm{Tr}_{C}\mathbf{\Gamma}_{\alpha}\left\{ 2\textrm{Re}\left(ie^{i\left(\omega-\mathbf{h}_{CC}^{\text{eff}}\right)\left(t-t_{0}\right)}\overline{e^{i\psi_{\alpha}\left(t,t_{0}\right)}}\mathbf{G}_{CC}^{r}\left(\omega\right)+\int_{t_{0}}^{t}d\tau e^{i\left(\omega-\mathbf{h}_{CC}^{\text{eff}}\right)\left(t-\tau\right)}\overline{e^{i\psi_{\alpha}\left(t,\tau\right)}}\right)\right.\nonumber \\
&-e^{-i\mathbf{h}_{CC}^{\text{eff}}\left(t-t_{0}\right)}\underset{\alpha'}{\sum}\left[\mathbf{G}_{CC}^{r}\left(\omega\right)\mathbf{\Gamma}_{\alpha'}\mathbf{G}_{CC}^{a}\left(\omega\right)+2\textrm{Re}\left(i\mathbf{G}_{CC}^{r}\left(\omega\right)\mathbf{\Gamma}_{\alpha'}\int_{t_{0}}^{t}d\tau e^{i\left(\omega-\mathbf{h}_{CC}^{\mathrm{eff}\dagger}\right)\left(\tau-t_{0}\right)}\overline{e^{i\psi_{\alpha}\left(\tau,t_{0}\right)}}\right)\right.\nonumber\\
&\left.\left.+\int_{t_{0}}^{t}d\tau\int_{t_{0}}^{t}d\bar{\tau}e^{i\left(\omega-\mathbf{h}_{CC}^{\text{eff}}\right)\left(\tau-t_{0}\right)}\mathbf{\Gamma}_{\alpha'}e^{i\left(\omega-\mathbf{h}_{CC}^{\mathrm{eff}\dagger}\right)\left(\bar{\tau}-t_{0}\right)}\overline{e^{-i\psi_{\alpha}\left(\tau,\bar{\tau}\right)}}\right]e^{i\mathbf{h}_{CC}^{\mathrm{eff}\dagger}\left(t-t_{0}\right)}\right\} ,
\end{align}
to give the full time-dependence of the current due to the stochastic and periodic fields. The complete expression for this is given as Eq. \eqref{eq:I_t} in Appendix \ref{Appendix_B}, where all frequency integrals are replaced with summations over Matsubara frequencies and we also expand in terms of the left and right eigenvectors of the effective Hamiltonian $\mathbf{h}_{CC}^{\text{eff}}$:
\begin{align}\label{eq:h_eff_LR}
    \mathbf{h}_{CC}^{\text{eff}}\left|\varphi_{i}^{R}\right\rangle =\bar{\varepsilon}_{i}\left|\varphi_{i}^{R}\right\rangle ;\thinspace\left\langle \varphi_{i}^{L}\right|\mathbf{h}_{CC}^{\text{eff}}=\bar{\varepsilon}_{i}\left\langle \varphi_{i}^{L}\right|.
\end{align}
Thus, the time-integrals appearing in Eq.~\eqref{eq:I_expand} can be carried out analytically, as demonstrated in Appendix~\ref{Appendix_B}.

The white noise case is recovered in the limit of zero correlation time $\tau_{c}\to 0$; in this limit only the $n=0$ term survives in the expansion~\eqref{eq:phase_factor} of the stochastic terms, which come in powers of the quantity $D/\omega_c=D\tau_{c}$. After taking the long time limit, transient modes are lost and we obtain the bias-averaged current:
%\textcolor{blue}{[LK: corrected an error in the eq below $z_{\alpha j} => y_{j,rs,m}^{(\alpha)}}$]}
%\begin{align}\label{eq:I_D_t}
\begin{align}\underset{t_{0}\rightarrow-\infty}{\lim}\overline{I_{\alpha}\left(t\right)} & \equiv I_{\alpha}\left(t;D,\omega_{c}\right)=\underset{rr'ss'}{\sum}\,\overset{\infty}{\underset{n=0}{\sum}}\,\overset{n}{\underset{m=0}{\sum}}\left(\frac{D}{\omega_{c}}\right)^{n}\frac{\left(-1\right)^{m}}{\left(n-m\right)!m!}\nonumber\\
 & \times\left\{ \frac{2}{\pi}\underset{j}{\sum}\textrm{Re}\left[i\frac{\left\langle \varphi_{j}^{L}\right|\mathbf{\Gamma}_{\alpha}\left|\varphi_{j}^{R}\right\rangle }{\left\langle \varphi_{j}^{L}\right|\left.\varphi_{j}^{R}\right\rangle }J_{rr'ss'}^{(\alpha)}e^{-i\left(r-r'\right)\phi_{\alpha}}e^{-i\Omega_{\alpha}\left(p_{1}\left(r-r'\right)+p_{2}\left(s-s'\right)\right)\left(t-t_{0}\right)}\right.\right.\nonumber\\
 & \left.\times\Psi\left(\frac{1}{2}-\frac{\beta}{2\pi i}y_{j,rs,m}^{(\alpha)}\right)\right]-\frac{1}{\pi}\underset{jk\alpha'}{\sum}\frac{\left\langle \varphi_{k}^{R}\right|\mathbf{\Gamma}_{\alpha}\left|\varphi_{j}^{R}\right\rangle \left\langle \varphi_{j}^{L}\right|\mathbf{\Gamma}_{\alpha'}\left|\varphi_{k}^{L}\right\rangle }{\left\langle \varphi_{j}^{L}\right|\left.\varphi_{j}^{R}\right\rangle \left\langle \varphi_{k}^{R}\right|\left.\varphi_{k}^{L}\right\rangle }J_{rr'ss'}^{(\alpha')}e^{-i\left(r-r'\right)\phi_{\alpha'}}\nonumber\\
 & \times\left.\frac{e^{-i\Omega_{\alpha'}\left(p_{1}\left(r-r'\right)+p_{2}\left(s-s'\right)\right)\left(t-t_{0}\right)}}{y_{j,rs,m}^{(\alpha)}-y_{k,r's',m}^{(\alpha)*}}\left[\Psi\left(\frac{1}{2}-\frac{\beta}{2\pi i}y_{j,rs,m}^{(\alpha)}\right)-\Psi\left(\frac{1}{2}+\frac{\beta}{2\pi i}y_{k,r's',m}^{(\alpha)*}\right)\right]\right\}\,, 
 \label{eq:I_D_t}
\end{align}
where
\begin{align}
y_{j,rs,m}^{(\alpha)}=\bar{\varepsilon}_{j}-\mu-V_{\alpha}-\Omega_{\alpha}\left(p_{1}r+p_{2}s\right)-i\left(D+m\omega_{c}\right)\,.\label{eq:y}
\end{align}
Here, we have introduced the digamma function as the logarithmic derivative of the gamma function, $\Psi\left(z\right)\equiv\frac{d\ln\Gamma\left(z\right)}{dz}$. 

For a consistency check, we may look at the static limit, $A_{\alpha}^{\left(1\right)}=0=A_{\alpha}^{\left(2\right)}$, for all $\alpha$. In this case, only the summation terms $r=r'=s=s'=0$ survive, since the Bessel function satisfies $J_{\nu}\left(0\right)=1$ for $\nu=0$ and $J_{\nu}\left(0\right)=0$ at all other natural  $\nu$. Using properties of the left/right eigenvectors, one can then obtain the adiabatic limit of Eq.~\eqref{eq:I_D_t}:
%\begin{align}\label{eq:I_D}
%I_{\alpha}\left(D,\omega_{c}\right)\equiv\underset{\forall %\alpha' 
\begin{align}%
I_{\alpha}\left(D,\omega_{c}\right)
\equiv\underset{A_{\alpha'}^{\left(1\right)},A_{\alpha'}^{\left(2\right)}\rightarrow0}{\lim}I_{\alpha}\left(t;D,\omega_{c}\right)=\frac{1}{\pi}\overset{\infty}{\underset{n=0}{\sum}}\,\overset{n}{\underset{m=0}{\sum}}\,\underset{jk}{\sum}\,\sum_{\alpha'\ne \alpha}\left(\frac{D}{\omega_{c}}\right)^{n}\frac{\left(-1\right)^{m}}{\left(n-m\right)!m!}\frac{\left\langle \varphi_{k}^{R}\right|\mathbf{\Gamma}_{\alpha}\left|\varphi_{j}^{R}\right\rangle \left\langle \varphi_{j}^{L}\right|\mathbf{\Gamma}_{\alpha'}\left|\varphi_{k}^{L}\right\rangle }{\left\langle \varphi_{j}^{L}\mid\varphi_{j}^{R}\right\rangle \left\langle \varphi_{k}^{R}\mid\varphi_{k}^{L}\right\rangle \left(\bar{\varepsilon}_{j}-\bar{\varepsilon}_{k}^{*}\right)}\nonumber\\
\times\left[\Psi\left(\frac{1}{2}-\frac{\beta}{2\pi i}y_{j,00,m}^{(\alpha)}\right)-\Psi\left(\frac{1}{2}+\frac{\beta}{2\pi i}y_{k,00,m}^{(\alpha)*}\right)-\Psi\left(\frac{1}{2}-\frac{\beta}{2\pi i}y_{j,00,m}^{(\alpha')}\right)+\Psi\left(\frac{1}{2}+\frac{\beta}{2\pi i}y_{k,00,m}^{(\alpha')*}\right)\right].
\label{eq:I_D}
\end{align}
This is none other than the derived expression (C6) in Ref.~\onlinecite{ridley_fluctuating-bias_2016} in the static bias case. We note that all time dependence has vanished. 
%{\textcolor{red}{[LK: in this case it is expected, but not in the general case I believe!]}} 
Eq. \eqref{eq:I_D} defines a \emph{quantum fluctuation-dissipation relation} for the nanojunction, as it relates conductance/resistance to temperature and fluctuation strength \cite{pottier2009nonequilibrium,ridley_fluctuating-bias_2016}.

Returning to Eq.~\eqref{eq:I_D_t}, we assume that the frequency of driving is lead-independent, $\Omega_{\alpha}=\Omega$ for all $\alpha$. This defines a driving period $\tau_{d}=2\pi/\Omega$ for the system. The  difference of the currents between any two leads is then time-averaged over the period of $\tau_{d}$ to get the noise-affected quantum pump current in the steady-state:
\begin{align}\label{eq:I_D_Pump}
    I_{\alpha\alpha'}^{\text{pump}}\equiv\underset{t_{0}\rightarrow-\infty}{\lim}\frac{\Omega}{2\pi}\int_{\tau}^{\tau+\frac{2\pi}{\Omega}}dt\left(I_{\alpha}\left(t;D,\omega_{c}\right)-I_{\alpha'}\left(t;D,\omega_{c}\right)\right),
\end{align}
where the $\tau$ appearing in the integral limits is an arbitrary time. We note that the net current between leads $\alpha$ and $\alpha'$ is computed here because it corresponds to the average DC quantity measured in a real experiment \cite{kohler2005driven}. It must also be emphasized that the `pump' terminology only applies to the case where the bias across the leads $\alpha, \alpha'$ is zero on average. Since the first moment of the noisy bias $V^{s}\left(t\right)$ is set to zero [Eq. \eqref{eq:1moment}], and the biharmonic terms $V_{\alpha}^{d}\left(t\right)$ are zero under the time average, this condition is satisfied when the constant shifts $V_{\alpha}$, $V_{\alpha'}$ are equal.

For simplicity, we specialize now to the two-lead terminal, and label left  and right leads by $L$ and $R$, respectively. We also assume that the amplitudes of driving are independent of the lead, $A_{L}^{\left(1\right)}=A^{\left(1\right)}=A_{R}^{\left(1\right)}$ and $A_{L}^{\left(2\right)}=A^{\left(2\right)}=A_{R}^{\left(2\right)}$. This leads to the stochastic bias-averaged quantum pump current
\begin{align}I_{LR}^{\text{pump}} & =\frac{1}{\pi}\underset{j,k}{\sum}\,\overset{\infty}{\underset{n=0}{\sum}}\,\overset{n}{\underset{m=0}{\sum}}\left(\frac{D}{\omega_{c}}\right)^{n}\frac{\left(-1\right)^{m}}{\left(n-m\right)!m!}\underset{r,r',s,s'}{\sum}\delta_{ss}^{rr'}\left(p_{i}\right)J_{rr'ss'}^{(\alpha)}\nonumber\\
 & \times\frac{\left\langle \varphi_{k}^{R}\right|\mathbf{\Gamma}_{L}\left|\varphi_{j}^{R}\right\rangle \left\langle \varphi_{j}^{L}\right|\mathbf{\Gamma}_{R}\left|\varphi_{k}^{L}\right\rangle +\left\langle \varphi_{k}^{R}\right|\mathbf{\Gamma}_{R}\left|\varphi_{j}^{R}\right\rangle \left\langle \varphi_{j}^{L}\right|\mathbf{\Gamma}_{L}\left|\varphi_{k}^{L}\right\rangle }{\left\langle \varphi_{j}^{L}\mid\varphi_{j}^{R}\right\rangle \left\langle \varphi_{k}^{R}\mid\varphi_{k}^{L}\right\rangle \left(\bar{\varepsilon}_{j}-\bar{\varepsilon}_{k}^{*}\right)}\nonumber\\
 & \times\left\{ e^{-i\left(r-r'\right)\phi_{L}}\left[\Psi\left(\frac{1}{2}-\frac{\beta}{2\pi i}y_{j,rs,m}^{(L)}\right)-\Psi\left(\frac{1}{2}+\frac{\beta}{2\pi i}y_{k,r's',m}^{\left(L\right)*}\right)\right]\right.\nonumber\\
 & \left.-e^{-i\left(r-r'\right)\phi_{R}}\left[\Psi\left(\frac{1}{2}-\frac{\beta}{2\pi i}y_{j,rs,m}^{(R)}\right)-\Psi\left(\frac{1}{2}+\frac{\beta}{2\pi i}y_{k,r's',m}^{\left(R\right)*}\right)\right]\right\} 
 \label{eq:ILR_pump}
\end{align}
where we have defined the modified Kronecker delta function:
\begin{equation}
\delta_{ss'}^{rr'}\left(p_{i}\right)\equiv\begin{cases}
1, & \textrm{if} \ p_{1}\left(r-r'\right)+p_{2}\left(s-s'\right)=0,\\
0, & \textrm{else}.
\end{cases}
\end{equation}

For evaluating Eq.~\eqref{eq:ILR_pump}, it is important to note that $A^{(1,2)}/\Omega$ grows in the slow-driving ($\Omega<A^{\left(1,2\right)}$) regime, approaching the adiabatic limit in Eq.~\eqref{eq:I_D} and may require a large number of Bessel function components for convergence.
%\mm{In the slow-driving regime, can we use Eq.(28) instead of Eq.(30)? I mean, in Eq.(28) we need to use the full time-dependent voltage instead of its DC part. Then Eq.(28) becomes time-dependent and we need average it over time to get its DC component.  In addition, can we add a line corresponding to Eq.(28) to Fig.6?}
In the white noise regime ($D < \omega_c$), bias fluctuations are small relative to the correlation frequency, so the stochastic-component summations converge fast. In the color noise regime ($D > \omega_c$), more stochastic components may be required to achieve convergence. 
%\textcolor{red}{If this paragraph is moved here (which is reasonable), then the absolute numbers should perhaps be removed as they relate to the specific parameter set in the next section. -rt}\textcolor{blue}{[LK: I have added above a clarification, would it be enough?]}

Formula \eqref{eq:ILR_pump} reduces to Eq. (23) in Ref.~\onlinecite{ridley_time-dependent_2017} in the limit of zero stochastic field ($n=0$) and equal biases $V_{L}=V_{R}$. It enables us to compute the interplay of the color noise with the external driving field. We finally note that charge is conserved across the nanojunction in the case of the driving bias considered here, i.e. 
\begin{equation}
    \underset{t_{0}\rightarrow-\infty}{\lim}\frac{\Omega}{2\pi}\int_{\tau}^{\tau+\frac{2\pi}{\Omega}}dt\left(I_{L}\left(t;D,\omega_{c}\right)+I_{R}\left(t;D,\omega_{c}\right)\right)=0,
\end{equation}
by the continuity Eq. \eqref{eq:continuity} and by the linearity of the bias-averaging, time averaging and time derivative operations.
\end{widetext}

\section{Results}\label{sec:numerics}

\subsection{Single-level molecular region}

In order to fix physical ideas, we first specialize the discussion to the case of a single level molecular region (the quantum dot case). In this case, the effective Hamiltonian is a scalar, $\mathbf{h}_{CC}^{\text{eff}}=\varepsilon_{0}-i\Gamma/2$ where $\varepsilon_{0}$ is the on-site dot energy and $\Gamma=\Gamma_{L}+\Gamma_{R}$ is the scalar level width for the two-lead model. We also set the chemical potential $\mu=0$ for simplicity of notation.

We consider the case, also considered in Ref.~\onlinecite{ridley_fluctuating-bias_2016} where there is no periodic driving. For simplicity, we also assume that the correlation frequency $\omega_{c}\to\infty$ (the white noise case), and we also take the zero temperature limit $\beta=1/k_{B}T\to\infty$. The bias is assumed to be applied symmetrically across the leads, such that $V_{L}=V=-V_{R}$. In this case, the bias-averaged current reduces to 
\begin{align}
I\left(V,D\right) & =\frac{2\Gamma_{L}\Gamma_{R}}{\pi\Gamma} \nonumber\\
&\times\left[\arctan\left(\frac{V-\varepsilon_{0}}{D+\Gamma}\right)+\arctan\left(\frac{V+\varepsilon_{0}}{D+\Gamma}\right)\right]
\end{align}
from which the differential conductance $G\left(V,D\right)=\frac{\partial I}{\partial V}$ can be extracted:
\begin{align}\label{eq:G_D_V}
&G\left(V,D\right)\nonumber \\
&=\frac{2\Gamma_{L}\Gamma_{R}\left(D+\Gamma\right)}{\pi\Gamma} \nonumber\\
&\times\left[\frac{1}{\left(V-\varepsilon_{0}\right)^2+\left(D+\Gamma\right)^2}  +\frac{1}{\left(V+\varepsilon_{0}\right)^2+\left(D+\Gamma\right)^2}\right]. 
\end{align}
Note that in the limit of large fluctuations, $D\gg V, \Gamma, \varepsilon_{0}$, the conductance $G$ becomes independent of the voltage $V$. This defines the ohmic regime~\cite{ridley_fluctuating-bias_2016}. From Eq.~\eqref{eq:G_D_V} we state a precise condition for stochastic enhancement of the conductance by the driving field, i.e. when the parameter set satisfies 
\begin{align}\label{eq:SR_Condition}
    G\left(V,D>0\right)>G\left(V,D=0\right)\,.
\end{align}
In the limit of large bias $V \gg D, \Gamma,\varepsilon_{0}$, condition Eq.~\eqref{eq:SR_Condition} is equivalent to $D+\Gamma>\Gamma$, which is always satisfied. However, in this latter regime, the conductance tends to zero. 

In general, 
%it seems we have to  
we should therefore
explore an intermediate parameter range $D \approx V$ in which there is a finite enhancement of a finite conductance due to the presence of the stochastic field.

\subsection{DQD molecular junction}
%\begin{itemize}
%\item Double quantum dot system coupled with two leads (WBLA)
%\item Energy scales: Bias window is ''small'' compared to the lead bandwidth (bias amplitude and fluctuations)
%\item Zero temperature / non-zero fluctuations
%\item Non-zero temperature / zero fluctuations
%\item Pump parameters: $p_1,p_2=1$, test the effect of $A^{(1)}, A^{(2}$, $\Omega_D$, $\phi_\alpha$
%\item Stochastic parameters $D$, $\omega_c\to\infty$ (white noise), $\omega_c\to 0$ (coloured noise)
%\end{itemize}

We now study numerically the double quantum dot (DQD) system shown in Fig.~\ref{DQD}, with onsite dot energies $\varepsilon_{1}$, $\varepsilon_{2}$, inter-dot coupling parameter $\lambda$ and coupling to the left and right leads described by the level broadening $\Gamma_{L}$, $\Gamma_{R}$ respectively, within the WBLA. The effective model Hamiltonian for the central molecular region is thus given by
\begin{equation}\label{eq:dotheff}
    \mathbf{h}_{CC}^{\text{eff}}=\left(\begin{array}{cc}
\varepsilon_{1}-i\Gamma_{L}/2 & \lambda\\
\lambda & \varepsilon_{2}-i\Gamma_{R}/2
\end{array}\right)\,,
\end{equation}
which determines the left/right eigenvalues/eigenvectors in Eq.~\eqref{eq:h_eff_LR}. It is seen that the state $\varepsilon_1$ is coupled to the left lead, while the state $\varepsilon_2$ to the right. Since we want to focus on the effect of the bias fluctuations, we simplify by considering the equal coupling to the leads, $\Gamma_L=\Gamma_R=\Gamma$, and then concentrate on the weak-coupling regime using $\Gamma=0.01$, where the WBLA is known to be accurate~\cite{verzijl_applicability_2013, covito_transient_2018}. Qualitatively, stronger coupling results in more broadened spectral features, cf.~Eq.~\eqref{eq:G_D_V}.

\begin{figure}[t]
%\centering
\includegraphics[width=0.475\textwidth]{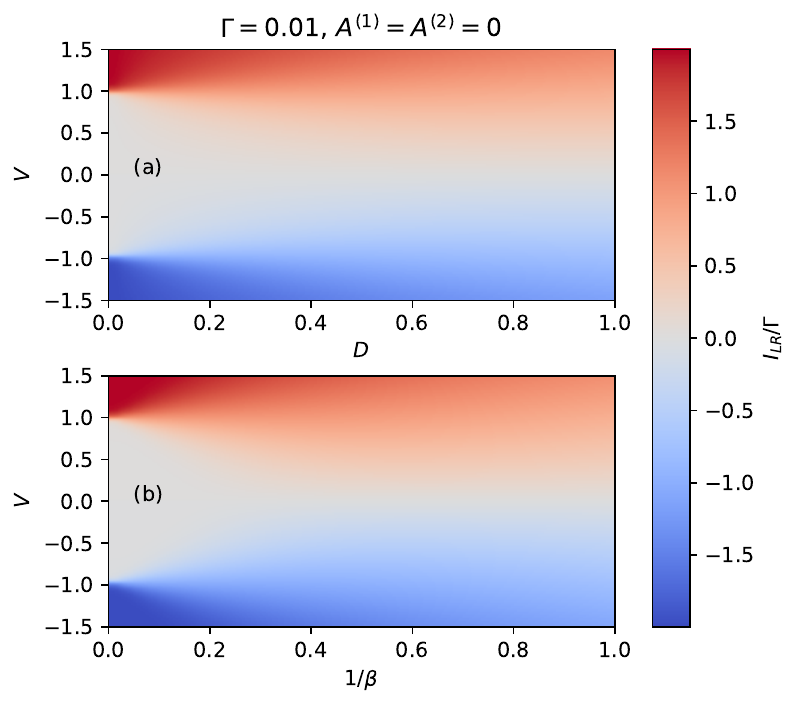}
\caption{Steady-state current-voltage characteristics of the statically driven DQD junction as a function of (a) $D$ at the (nearly) zero-temperature limit ($\beta=100$) and (b) $k_{\mathrm{B}}T$ at the zero-fluctuation limit ($D=0$).}
\label{fig:stochcurrentGamma0.01}
\end{figure}

\subsection{Stochastic field with constant driving}\label{sec:constant_V}
    
We first consider the case with no time-periodic driving term in the lead bias and $V_{L}=-V_{R} = V$,
%\mm{V or V/2 ?}
%\textcolor{red}{should this be V/2??}
%\textcolor{blue}{I think it is $V$, because the peaks in Fig.~\ref{fig:stochcurrentGamma0.01} appear at $V=\pm 1$ which are the resonant levels of the DQD. We could maybe change this definition accordingly in Eq.~\eqref{eq:G_D_V}.-Riku}
%\textcolor{red}{OK - Mike}
in the white noise case and with the DQD Hamiltonian parameterized by $\varepsilon_{1}=0=\varepsilon_{2}$ and $\lambda=-1$. Fig.~\ref{fig:stochcurrentGamma0.01}(a) displays the steady-state current-voltage characteristics as a function of the stochastic parameter $D$, at zero temperature. We see a transition from the step-like current-voltage characteristics to a linear, ohmic dependence of the current on the voltage with increasing $D$. This is compared to the case of zero fluctuations and varying temperature in Fig.~\ref{fig:stochcurrentGamma0.01}(b), from which it is apparent that the effect of the environmental fluctuations on the current is qualitatively identical to that of temperature. We note that this is not true for all variables: the temperature and the environmental fluctuations can have qualitatively different effects on the current noise~\cite{Moskalets_2016}. We also note that, for this case of a static bias plus white noise, the results of computing Eq.~\eqref{eq:ILR_pump} are identical to those obtained from the adiabatic limit 
%$\underset{\forall \gamma A_{\gamma}^{\left(1\right)},A_{\gamma}^{\left(2\right)} \rightarrow 0}{\lim}I_{\alpha}\left(t;D,\omega_{c}\right)$ 
in Eq.~\eqref{eq:I_D}, because in this limit the bias-averaged current exhibits no time-dependence.

%\begin{figure}[htp]
%\centering
%    \includegraphics[clip, width=\linewidth]{wire-stochcurrentGamma0.1.pdf}
%    \caption{Steady-state $I-V$ characteristic in the DQD as a function of $D$, at $k_BT=0$ and $\Gamma=0.1$.\textcolor{red}{TO BE REPLACED WITH A COMBINED FIGURE}} 
%    \label{wire-stochcurrentGamma0.1}
%\end{figure}

%\begin{figure}[htp]
%\centering
%    \includegraphics[clip, width=\linewidth]{wire-stochcurrentGamma0.1beta.pdf}
%    \caption{Steady-state $I-V$ characteristic in the DQD as a function of $k_BT$, at $D=0$  and $\Gamma=0.1$.\textcolor{red}{TO BE REPLACED WITH A COMBINED FIGURE}} 
%    \label{wire-stochcurrentGamma0.1beta}
%\end{figure}

\begin{figure}[t]
%\centering
\includegraphics[width=0.45\textwidth]{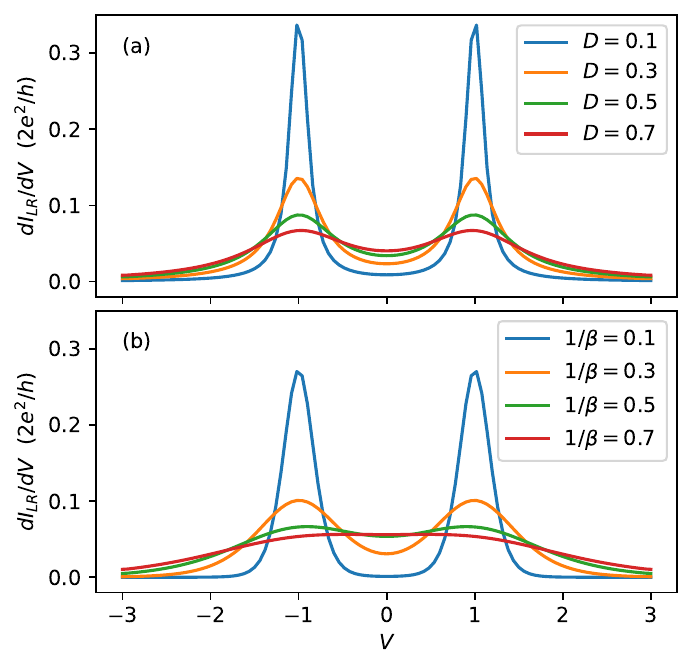}
\caption{Differential conductance for specific values of (a) fluctuation strength $D$ at the zero-temperature limit ($\beta=100$), and (b) temperature $1/\beta$ at the zero-fluctuation limit ($D=0$), corresponding to Fig.~\ref{fig:stochcurrentGamma0.01}.}
\label{fig:figwire-IVGamma0.01}
\end{figure}

In Fig.~\ref{fig:figwire-IVGamma0.01}, we show differential conductances $G(V,D)$ in units of the conductance quantum $2e^2/h$, evaluated by numerical differentiation from the pump currents shown in Fig.~\ref{fig:stochcurrentGamma0.01}, for different values of the fluctuation strength $D$ (panel a) and the temperature $1/\beta$ (panel b). Similar to the discussion of Fig.~\ref{fig:stochcurrentGamma0.01}, we observe that the typical double-peak structure in the conductance is progressively smeared out as the fluctuation strength and temperature are increased.

%\begin{figure}[htp]
%\centering
%    \includegraphics[clip, width=\linewidth]{wire-IVGamma0.01.pdf}
%    \caption{Current-voltage characteristics and corresponding conductance for $\Gamma=0.01$ and varying $D$.\textcolor{red}{TO BE REPLACED WITH A MORE SMOOTH DATA SET}}
%    \label{fig:figwire-IVGamma0.01}
%\end{figure}

\subsection{Stochastic field with periodic driving}\label{sec:periodic_V}
%We now move on to the case of the driven DQD. Fig. \ref{fig:Resonance} shows the pump current $I_{LR}^{pump}$ as a function of the phase difference $\phi=\phi_{L}-\phi_{R}$ and the fluctuation strength $D$, for different values of the static bias $V_{L}=V=V_{R}$. In these calculations, we set $p_{1}=1=p_{2}$ and choose the driving amplitudes to be half the voltage shift, $A^{(1)}=V/2=A^{(2)}$. Interestingly, whereas in the $V=0.5$ and $V=1.5$ cases there is a rapid decay of the pump current with fluctuation strength, in the intermediate regime of $V=1$ we observe an initial decay followed by a resonant peak in the pump current at $D\approx1=V$.\textcolor{red}{discuss the fact that SR seems to be activated by the sinusoidal bias, but isn't obviously present in the static bias case. In this sense the SR in all calculations that follow is truly photon-assisted. Schematic diagram?}

\begin{figure}[t]
\includegraphics[width=0.45\textwidth]{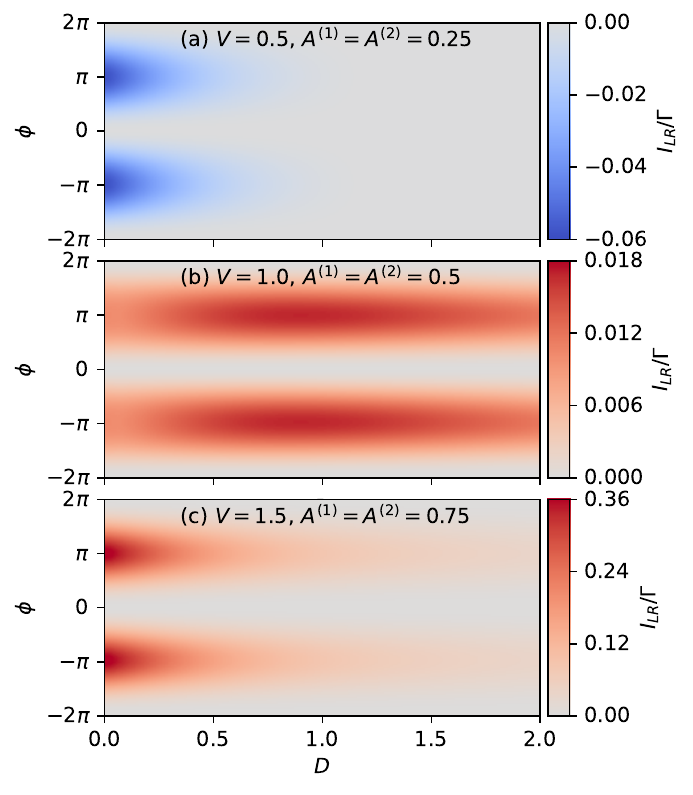}
\caption{AC driven DQD junction: pump current (colormap) in terms of the total phase difference $\phi=\phi_L-\phi_R$ (vertical axes) and the bias fluctuation strength $D$ (horizontal axes) for different values of the static bias voltage (a) $V=0.5$, (b) $V=1.0$, and (c) $V=1.5$.}
\label{fig:driven}
\end{figure}

We now turn to the case of the periodically driven DQD. Fig.~\ref{fig:driven} illustrates the pump current as a function of the phase difference $\phi = \phi_L - \phi_R$ and the fluctuation strength $D$, for different values of the static bias $V_L = V_R = V$, 
%\textcolor{red}{should it be V/2??}
see Eqs.~\eqref{eq:driving} and~\eqref{eq:drivingprotocol}: to better isolate the behavior of the stochastic drive, we fix the constant value of the bias to be equal on both left and right leads, so that applying a phase difference $\phi=\pm\pi$ corresponds to the entirely antisymmetric bias window. Also, to simplify the large parameter space of the driving protocol, we set $p_1 = p_2 = 1$, take the modulating amplitudes as half the voltage shift, $A^{(1)} = A^{(2)} = V/2$, and set first the driving frequency $\Omega=1$. 

Interestingly, in the $V = 0.5$ and $V = 1.5$ cases [Figs.~\ref{fig:driven}(a,c)], the pump current decays rapidly with increasing fluctuation strength. However, in the intermediate regime $V = 1.0$ [Fig.~\ref{fig:driven}(b)], we observe a clear stochastic enhancement of the pump current near $D = 1$. At zero fluctuations, a sizable pump current exists at the off-phase points $\phi = \pm\pi$ in all cases. This arises because, even if the constant bias does not align with the resonant energy levels (the real parts of the eigenvalues $\bar{\varepsilon}_{1,2}$  of the matrix in Eq.~\eqref{eq:dotheff}) of the DQD at $\mathrm{Re} \, \bar{\varepsilon}_{1,2} = \mp 1$, the modulating amplitudes generate a directed current~\cite{ridley_time-dependent_2017}. (Similarly, the in-phase points $\phi=0,\pm2\pi$ show diminishing pump currents in all cases.) However, the rectification effect is suppressed by bias fluctuations unless the constant bias aligns with the resonant levels ($V\approx 1.0$). When alignment occurs, fluctuations enhance rectification, leading to photon-assisted SR.

In Fig.~\ref{fig:driven}, we also find that the pump current changes sign as the applied bias crosses the value related to the resonant energy levels of the DQD system  [Figs.~\ref{fig:driven}(a,c)]. This effect can be understood by photon-assisted electron and hole transfer processes~\cite{ridley_time-dependent_2017, ridley2022many}. The chemical potential of the coupled system is set in the middle of the DQD energy levels ($\mu=0$) indicating the corresponding electron populations in the leads. The effect is then seen as a directed current through the DQD from left to right (positive current) or opposite direction (negative current).

\begin{figure}[t]
\includegraphics[width=0.5\textwidth]{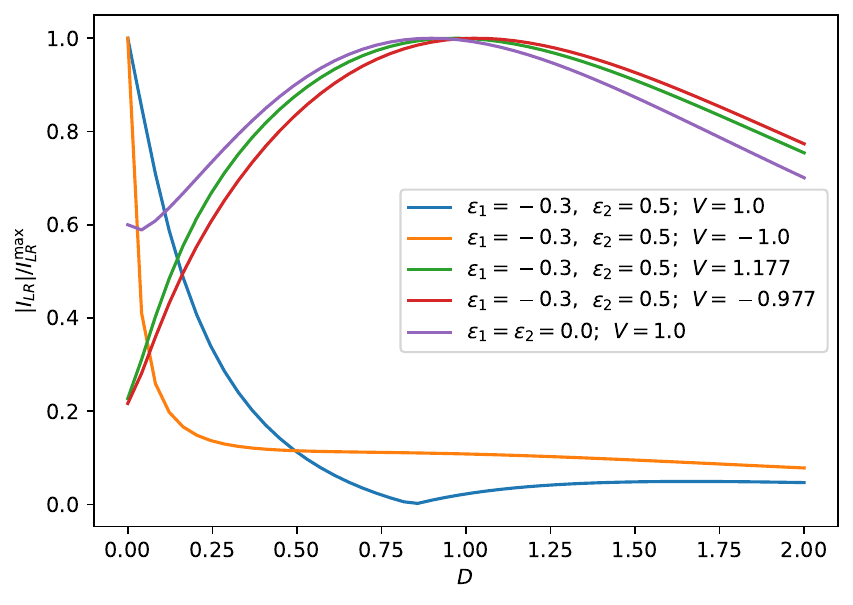}
\caption{Absolute value of the pump current, normalized by its maximum value, in terms of the fluctuation strength $D$ in the white noise case ($\omega_c=10$) when the DQD is subjected to on-site potentials $\varepsilon_{1,2}$ specified in the legend.}
\label{fig:onsite}
\end{figure}

The parameter space of the driving protocol in Eq.~\eqref{eq:driving} is very large. Since we are restricting the discussion on specific parameters, it is possible that there are more optimized settings for enhancing the SR effect even more. For this reason, we now refer to the situation in Fig.~\ref{fig:driven}(b), $V = 1.0$ and $\phi = \pi$, as the `optimum SR' scenario (within our parameter regime). In this situation,
%Focusing on the maximum SR scenario, $V = 1.0$ and $\phi = \pi$, 
we analyze the impact of other parameters in the DQD junction. Fig.~\ref{fig:onsite} illustrates the effect of on-site potentials $\varepsilon_{1,2}$ in Eq.~\eqref{eq:dotheff}.
%{\textcolor{red}{[LK: as you talk about the "effect" of the energy levels, I think you also need to provide a curve corresponding to the previous case of them being equal and 0, for comparison.]}}
These parameters can be tuned in practice using local gate voltages~\cite{hanson_spins_2007}, which lift particle-hole degeneracy and shift the DQD resonant levels. For this, we set $\varepsilon_1=-0.3$ and $\varepsilon_2=0.5$ which modifies the energy levels of ${\mathbf h}^{\rm eff}_{CC}$ from $\mathrm{Re} \, \bar{\varepsilon}_{1,2} = \mp 1$ to $\mathrm{Re} \, \bar{\varepsilon}_{1,2} = \{-0{.}977,1{.}177\}$.
%{\textcolor{red}{[LK: where were the energy levels $\pm1$? you have $V$ in Fig. 5 being either $\pm 1$ and 1.177 and -0.977]}} 
Applying the same static bias voltage profile, $V=1.0$, that induces strong SR in Fig.~\ref{fig:driven}, now results in a decaying behavior with increasing bias fluctuations. Note, however, the currents are not necessarily symmetric when mirroring the bias due to the broken particle-hole symmetry and the chemical potential being kept fixed at $\mu=0$.
%{\textcolor{red}{[LK: is this obvious? The leads are identical, why the current should be different?] I;d understand that if we went beyond WBLA and used the actual DOS below and above Fermi energy, but within the WBLA?]}} 
Adjusting the constant bias to the modified resonant levels, $V=-0{.}977$ or $V=1{.}177$,
%\mm{Can we specify  the adjustment?: \( V=\dots\)}
restores the SR effect, demonstrating the robustness of SR against local fields that break particle-hole symmetry.

\begin{figure}[t]
    \centering
    \includegraphics[width=0.45\textwidth]{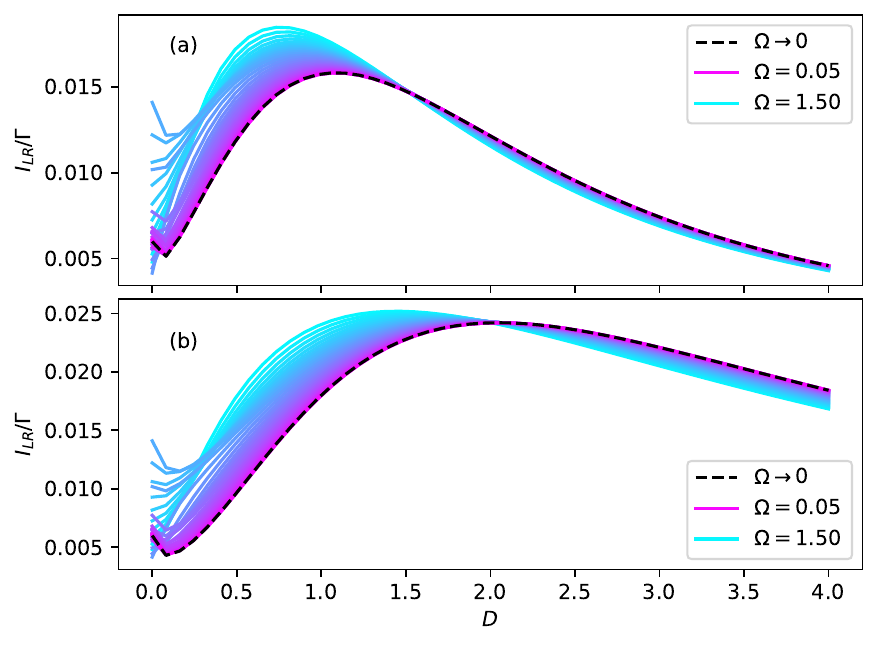}
    \caption{Pump current in (a) white noise, $\omega_c=10$, and (b) color noise, $\omega_c=1$, regimes in terms of the fluctuation strength ($D$) for several driving frequencies between $\Omega=0.05$ (magenta) and $\Omega=1.5$ (cyan) with steps of $0.05$. The black dashed lines display the adiabatic limit according to Eq.~\eqref{eq:I_D}.
    %\mm{Interesting, why in panel (b) at D=2 the result seems to be independent of frequency but it is still different from the adiabatic one? Why is there a crossover near D=2 from the region where the current increases with increasing frequency to the region where the current decreases with frequency?  Apart from this, it is amazing how good the adiabatic limit reproduces the low-frequency result (well, better to say, the adiabatic equation provides a lower bound to this effect)}
    %\textcolor{green}{rt: In both panels, there seems to be a point in $D$ where all lines overlap, but it is not exactly matched by the adiabatic limit. This is probably due to the inaccuracy of the procedure to calculate the adiabatic limit. We selected $11$ points for $V_\alpha$ along half a `period' of oscillation, which were then averaged over. It may be that a higher number of points would make the dashed line collapse on the color lines. Around the point where the lines meet (in $D$) I do not have an explanation why the value of pump current increases or decreases with driving frequency. Maybe some timescale argument between the driving and the correlation time?}
    %\mm{In my view, the accuracy of adiabatic calculations can be made much higher compared to the accuracy of non-adiabatic calculations. As a result, the adiabatic calculations can act as a benchmark for low-frequency calculations.}
    }
    \label{fig:comparison-noises}
\end{figure}

The position of the photon-assisted SR peak with respect to fluctuation strength $D$ depends on the driving frequency. Fig.~\ref{fig:comparison-noises}(a) shows a frequency sweep $\Omega \in [0.05, 1.5]$ within the white-noise regime ($\omega_c = 10$), still focusing on the optimal SR case $V = 1.0$ and $\phi = \pi$. (Also, the on-site potentials are again set to  $\varepsilon_1=\varepsilon_2=0$.)
%{\textcolor{red}{[LK: is it obvious that the maximum SR would always be at these values and will not shift if we change say the positions of the energy levels and/or $\Omega$? If that is not the case (as I'd believe), then may be it makes sense to locate the max of SR for various parameters? A "phase" diagram?]}}
The SR peak shifts with driving frequency: a faster driving ($\Omega=1.5$) causes SR at weaker fluctuations ($D=0.7$), slower driving ($\Omega=0.05$) requires stronger fluctuations ($D=1.2$) for SR, but eventually, all curves decay at large $D$. We also observed (not shown) that beyond a threshold in increasing driving frequencies ($\Omega \gtrsim 2$), the system cannot respond fast enough, leading to monotonically decreasing current with $D$. Fig.~\ref{fig:comparison-noises}(b) shows the color-noise regime ($\omega_c = 1$), where the SR peak shifts to higher $D$ for both fast and slow driving. As $\omega_c$ decreases, the correlation time $\tau_c = 1/\omega_c$ increases, resulting in a `smearing' out of the SR peak due to an increased effective range of resonant bias values. The effect of the color noise is therefore to improve the stability of the transport signal against bias fluctuations. 

%%Considering the definition of current as product between the current density and the surface section through which the electrons goes, it is possible to underline three physical quantities: the volumetric density of electron, the speed od these electrons and the above-mentioned surface. 

%We could interpret the  resultant current, due to fluctuation, in several ways that could be superimposed effects of the behaviour of the three physical quantities. Greater fluctuations could have electrons reach a certain threshold that once it has been overcame the current decays because the applied fluctuation is so high that it does not respect the resonance value. This could be combined with a vision of a physical channel, that according to the value of fluctuation, has been made larger or smaller for electrons to go through and this implies a higher or lower value of current.

%%Still we could consider a comparison in some way similar to localized surface plasmonic resonance or also more simply to a dumped and forced harmonic oscillator in which there are some characterizing parameters that establish several distinguishable regimes according to their value with respect the dimensions of the objects involved in the phenomenon (in the case of LSPR they are the wavelength of the incident electric field and the radius of the metallic nanoparticle)

Related to the convergence discussion after Eq.~\eqref{eq:ILR_pump}, here the most challenging case ($\Omega=0.05$, $\omega_c=1.0$, $D=4.0$) requires $\{|r|, |r'|, |s|, |s'|\}_{\mathrm{max}} \gtrsim 17$ Bessel function components and $\{n, m\}_{\mathrm{max}} \gtrsim 13$ stochastic components for convergence. This calculation then involves about $6\cdot 10^7$ calls to the GSL special function library~\cite{gsl}.

The slow-driving regime $\Omega\to 0$   can also be approached by employing the adiabatic limit in Eq.~\eqref{eq:I_D}. In place of the constant shifts $V_\alpha$, $V_{\alpha'}$, we insert a set of `time-dependent' values from Eq.~\eqref{eq:driving} into Eq. ~\eqref{eq:I_D}, and then average over one period of oscillation. 
While this procedure is not as general as the bias averaging carried out with Eq.~\eqref{eq:phase_factor}, it is expected to provide the adiabatic limit, \(\Omega\to 0\), for the AC-driven pump currents in the absence of a stochastic time-dependent fields, \(V_\alpha^{s}(t)=0\),
without having to numerically converge a large number of Bessel function components~\cite{moskalets2002floquet, moskalets2011scattering}. 
This procedure happens to work in the presence of a stochastic noise as well.
The result is shown as black dashed lines in Fig.~\ref{fig:comparison-noises}. Importantly, we see the SR effect is clearly visible also in this case, practically coinciding with the $\Omega=0.05$ cases. While the peak position changes with $\Omega$, the overall SR effect is not limited by this parameter, and could be detectable with state-of-the-art electronics operating at the terahertz regime~\cite{urteaga_inp_2017}.

\begin{figure}[t]
\includegraphics[width=0.45\textwidth]{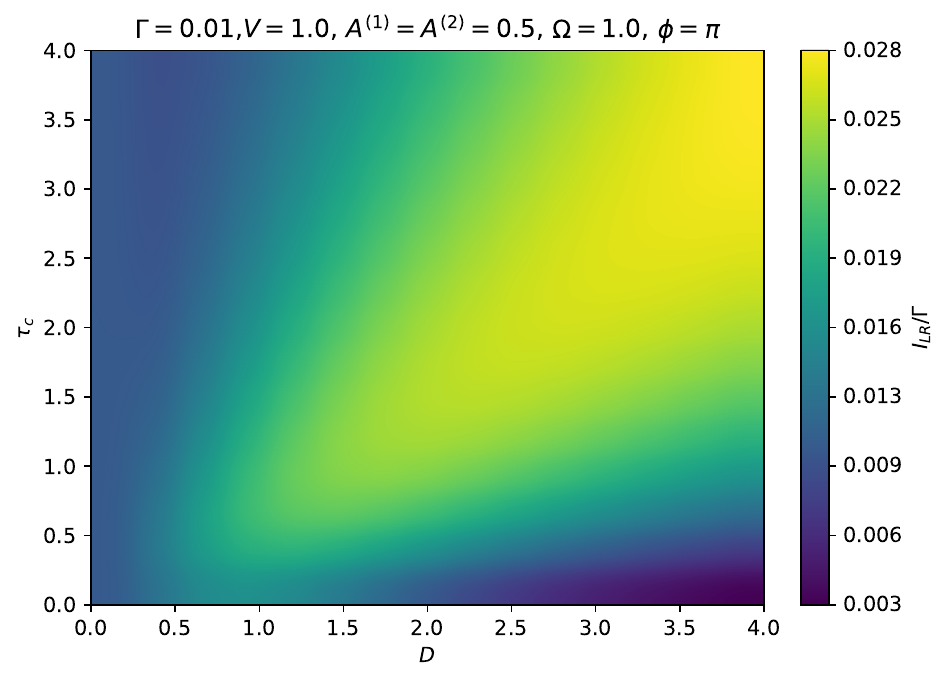}
\caption{ Stochastic pump current (colormap) as function of the inverse of the correlation frequency $\tau_c=1/\omega_c$ (vertical axes) and the strength of bias fluctuation $D$ (horizontal axes).}
\label{fig:corrtime}
\end{figure}

The shift of the SR peak between white- and color-noise regimes is further analyzed in Fig.~\ref{fig:corrtime} by varying both correlation time $\tau_c = 1/\omega_c$ and the fluctuation strength $D$ while keeping the driving frequency fixed at $\Omega=1.0$. In addition, this calculation again focuses on the optimum SR case with $V = 1.0$ and $\phi = \pi$. The pump current maximum moves roughly diagonally in the $(D, \tau_c)$-plane but becomes smeared at large $(D,\tau_c)$ as the DQD system operates through stronger bias fluctuations and color noise.

\section{Conclusions}\label{sec:concl}

We have shown in this work that a nanojunction subject to a noisy external field can make \emph{positive} use of the noise when assisted by a periodically-varying driving voltage. Furthermore, we have mapped out the parameter range in which this stochastic resonance effect is optimized for a double quantum dot nanojunction. To this end, 
%we have developed a novel formalism, based on the TD-LB method, 
we have generalized the time-dependent Landauer-B{\"u}ttiker method
to incorporate both stochastic and deterministic terms in the bias profile and have shown how the resulting non-linear time-dependent current response reduces to 
earlier results reported in the literature.
%published results in the literature. 
This enabled us to optimize the stochastic resonance effect in the pumped electron current, and to demonstrate its dependence on intramolecular frequencies. In addition, we have computed the effect of varying the driving frequency and the fluctuation correlation time on SR, demonstrating its resilience across a wide range of parameter regimes.
While we established how the SR depends on the frequency and fluctuations of the bias, we also confirmed the overall effect is present at the adiabatic limit, and could be observed with terahertz-scale electronics~\cite{urteaga_inp_2017}.

In future work, we will investigate SR in 
%larger molecular structures 
more structurally diverse nanojunctions
such as graphene nanoribbons, for which the validity of the TD-LB formalism is well-established \cite{tuovinen_time-dependent_2014, gomes_da_rocha_curvature_2015, ridley_time-dependent_2017}. We will also study the complex interplay between the timescales of the stochastic driving field, dissipation and driving time period, using Eqs. \eqref{eq:I_D_t} and \eqref{eq:I_t}. The formalism developed here may also be extended to study SR in energy currents \cite{covito_transient_2018}, which will enable us to harness it for the optimization of thermoelectric effects. Finally, we may consider the effect of the noisy bias on the current-current correlations and quantum noise, using the formalism developed in Ref.~\onlinecite{ridley_partition-free_2017}. This will enable us to look at the effects of, e.g., noise on the electronic traversal time \cite{ridley_electron_2019} in the context of transiently emerging topological phenomena in nanojunctions out of equilibrium~\cite{tuovinen_distinguishing_2019,baranski_dynamical_2021,tuovinen_electron_2021,gorski_nonlocal_2024}.

\begin{acknowledgments} 
M.R. acknowledges support from the European Union’s Horizon Europe research and innovation programme under grant agreement No. 101178170. M.M. acknowledges the support from CSIC/IUCRAN2022 under Grant No. UCRAN20029. L.B. and R.T. acknowledge the EffQSim project funded by the Jane and Aatos Erkko Foundation. We also acknowledge grants of computer capacity from the Finnish Grid and Cloud Infrastructure (persistent identifier urn:nbn:fi:research-infras-2016072533).
\end{acknowledgments}
    
\appendix
\begin{widetext}
\section{Green's function and self-energy components in the WBLA}\label{sec:Appendix_A}

In this section, we first list the Green's function components for the generic time-dependent Hamiltonian studied in this paper [Eq. \eqref{eq:Hamiltonian}]. These have been derived in previous works on the TD-LB formalism \cite{stefanucci_nonequilibrium_2013,ridley_current_2015,ridley_fluctuating-bias_2016}. In the arguments of these functions, we use the notation $t$ to denote real times taken on the horizontal branch of the Konstantinov-Perel' contour, and $\tau$ denotes the imaginary part of $t_0-i\tau$ taken on the Matsubara branch. The real-time components read:
\begin{align}\label{eq:GF_r}
    \mathbf{G}_{CC}^{r}\left(t_{1},t_{2}\right) = -i\theta\left(t_{1}-t_{2}\right)e^{-i\mathbf{h}{}_{CC}^{\text{eff}}\left(t_{1}-t_{2}\right)} \,,
\end{align}
\begin{align}\label{eq:GF_a}
    \mathbf{G}_{CC}^{a}\left(t_{1},t_{2}\right) = i\theta\left(t_{2}-t_{1}\right)e^{-i\left(\mathbf{h}{}_{CC}^{\text{eff}}\right)^{\dagger}\left(t_{1}-t_{2}\right)} \,,
\end{align}
\begin{align}\label{eq:GF_lessgtr}
    \mathbf{G}_{CC}^{\lessgtr}\left( t_{1},t_{2} \right)=\pm i\int\frac{d\omega}{2\pi}f\left(\pm\left(\omega-\mu\right)\right)\underset{\beta}{\sum}\mathbf{S}_{\beta}\left(t_{1},t_{0};\omega\right)\Gamma_{\beta}\mathbf{S}_{\beta}^{\dagger}\left(t_{2},t_{0};\omega\right)\,.
\end{align}
Here, the matrix ${\bf S}_{\beta}$ is defined in Eq.~\eqref{eq:S_alpha}. The mixed and Matsubara components read:
\begin{equation}\label{eq:GF_left}
\mathbf{G}_{CC}^{\rceil}\left(t_{1},\tau_{2}\right)=e^{-i\mathbf{h}_{CC}^{\text{eff}}\left(t_{1}-t_{0}\right)}\left[\mathbf{G}_{CC}^{M}\left(0^{+},\tau_{2}\right)-i\underset{t_0}{\overset{t_1}{\int}}d\bar{t}e^{i\mathbf{h}_{CC}^{\text{eff}}\left(\bar{t}-t_{0}\right)}\left(\mathbf{\Sigma}_{em}^{\rceil}\star\mathbf{G}_{CC}^{M}\right)\left(\bar{t},\tau_{2}\right)\right],\,
\end{equation}
\begin{equation}\label{eq:GF_right}
\mathbf{G}_{CC}^{\lceil}\left(\tau_{1},t_{2}\right)=\left[\mathbf{G}_{CC}^{M}\left(\tau_{1},0^{+}\right)+i\underset{t_0}{\overset{t_2}{\int}}d\bar{t}\left(\mathbf{G}_{CC}^{M}\star\mathbf{\Sigma}_{em}^{\lceil}\right)\left(\tau_{1},\bar{t}\right)e^{-i\left(\mathbf{h}_{CC}^{\text{eff}}\right)^{\dagger}\left(\bar{t}-t_{0}\right)}\right]e^{i\left(\mathbf{h}_{CC}^{\text{eff}}\right)^{\dagger}\left(t_{2}-t_{0}\right)} \,,
\end{equation}
\begin{equation}\label{eq:GF_M}
\mathbf{G}_{CC}^{M}\left(\tau_{1},\tau_{2}\right)=\frac{i}{\beta}\sum_{q}e^{-\omega_{q}\left(\tau_{1}-\tau_{2}\right)}\mathbf{G}_{CC}^{M}\left(\omega_{q}\right)\,.
\end{equation}
Here we have defined the Matsubara frequencies $\omega_{q}=i\left(2q+1\right)\pi/\beta$, and the Fourier transformed Matsubara Green's function 
\begin{align}\label{eq:GF_M_Omega}
    \mathbf{G}_{CC}^{M}\left(\omega_{q}\right)=\begin{cases}
    \left[\left(\omega_{q}+\mu\right)\mathbf{1}_{CC}-\mathbf{h}_{CC}^{\text{eff}}\right]^{-1}, & \text{Im}\left(\omega_{q}\right)>0,\\
    \left[\left(\omega_{q}+\mu\right)\mathbf{1}_{CC}-\left(\mathbf{h}_{CC}^{\text{eff}}\right)^\dagger\right]^{-1}, & \text{Im}\left(\omega_{q}\right)<0.
    \end{cases}
\end{align}

The corresponding embedding self-energy components are as follows:
\begin{align}
    \mathbf{\Sigma}_{em}^{r/a}\left(t_{1},t_{2}\right)=\mp\frac{i}{2}\mathbf{\Gamma}\delta\left(t_{1}-t_{2}\right),
\end{align}
\begin{align}
    \mathbf{\Sigma}_{em}^{\lessgtr}\left(t_{1},t_{2}\right)=\pm i\underset{\alpha}{\sum}\mathbf{\Gamma}_{\alpha}e^{-i\psi_{\alpha}\left(t_{1},t_{2}\right)}\int\frac{d\omega}{2\pi}f\left(\pm\left(\omega-\mu\right)\right)e^{-i\omega\left(t_{1}-t_{2}\right)},
\end{align}
\begin{align}
    \mathbf{\Sigma}_{em}^{\rceil}\left(t_{1},\tau_{2}\right)=\frac{i}{\beta}\underset{\alpha}{\sum}\mathbf{\Gamma}_{\alpha}e^{-i\psi_{\alpha}\left(t_{1},t_{0}\right)}\underset{q}{\sum}e^{\omega_{q}\tau_{2}}\int\frac{d\omega}{2\pi}\frac{e^{-i\omega\left(t_{1}-t_{0}\right)}}{\omega_{q}-\omega+\mu},
\end{align}
\begin{align}
    \mathbf{\Sigma}_{em}^{\lceil}\left(\tau_{1},t_{2}\right)=\frac{i}{\beta}\underset{\alpha}{\sum}\mathbf{\Gamma}_{\alpha}e^{i\psi_{\alpha}\left(t_{2},t_{0}\right)}\underset{q}{\sum}e^{-\omega_{q}\tau_{1}}\int\frac{d\omega}{2\pi}\frac{e^{i\omega\left(t_{2}-t_{0}\right)}}{\omega_{q}-\omega+\mu},
\end{align}
\begin{align}
    \mathbf{\Sigma}_{em}^{M}\left(\tau_{1},\tau_{2}\right)=-\frac{1}{2\beta}\mathbf{\Gamma}\underset{q}{\sum}\xi_{q}e^{-\omega_{q}\left(\tau_{1}-\tau_{2}\right)}.
\end{align}
In the latter expression, $\xi_q=+1$ when $\textrm{Im}\left(\omega_{q}\right)>0$ and $\xi_q=-1$ when $\textrm{Im}\left(\omega_{q}\right)<0$.

\section{Full expansion of the bias-averaged current}\label{Appendix_B}

The numerical results in this paper are obtained by analytically removing all time and frequency integrals in Eq.~\eqref{eq:I_expand}. The time integrals are removed by replacing all phase factors with the expansion in Eq.~\eqref{eq:phase_factor}. The frequency integrals are removed by making use of the following simple pole expansion of the Fermi function:
\begin{align}\label{eq:Fermi}
    f\left(x\right)=\frac{1}{e^{\beta x}+1}=\frac{1}{2}-\overset{\infty}{\underset{n=1}{\sum}}\left(\frac{1}{\beta x+i\zeta_{n}}+\frac{1}{\beta x-i\zeta_{n}}\right),
\end{align}
where the $\zeta_{n}=\pi\left(2n-1\right)$ are the so-called Matsubara frequencies \cite{ridley_time-dependent_2017}. When Eq. \eqref{eq:Fermi} is inserted into Eq. \eqref{eq:I_expand}, the resulting expression is given in terms of the digamma function, and the Hurwitz-Lerch transcendental function \cite{hurwitz1887,lerch1887}, which is defined as follows:
\begin{align}\label{eq:HLT}
    \Phi\left(z,s,a\right)\equiv\overset{\infty}{\underset{n=0}{\sum}}\frac{z^{n}}{\left(n+a\right)^{s}}.
\end{align}
For notational convenience, we also introduce the following compact object in terms of Eq. \eqref{eq:HLT}:
\begin{align}\label{eq:Phi_overline}    
    \overline{\Phi}\left(\tau,\beta,z\right)\equiv\exp\left(-\frac{\pi\tau}{\beta}\right)\Phi\left(e^{-\frac{2\pi\tau}{\beta}},1,\frac{1}{2}+\frac{\beta z}{2\pi i}\right).
\end{align}

The expression which results from the removal of all integrals, although cumbersome, enables one to evaluate the current as a `single shot' function of time:
\begin{align}\overline{I_{\alpha}\left(t\right)} & =\sum_{j}\left[\textrm{Re}\frac{2i}{\pi}\frac{\left\langle \varphi_{j}^{L}\right|\Gamma_{\alpha}\left|\varphi_{j}^{R}\right\rangle }{\left\langle \varphi_{j}^{L}\right|\left.\varphi_{j}^{R}\right\rangle }K_{\alpha j}(t)-\underset{k}{\sum}\underset{\alpha'}{\sum}\frac{1}{\pi}\frac{\left\langle \varphi_{k}^{R}\right|\Gamma_{\alpha}\left|\varphi_{j}^{R}\right\rangle \left\langle \varphi_{j}^{L}\right|\Gamma_{\alpha'}\left|\varphi_{k}^{L}\right\rangle }{\left\langle \varphi_{j}^{L}\right|\left.\varphi_{j}^{R}\right\rangle \left\langle \varphi_{k}^{R}\right|\left.\varphi_{k}^{L}\right\rangle }P_{\alpha\alpha'jk}(t)\right]\,,\label{eq:I_t}
\end{align}
where 
\begin{align}
K_{\alpha j}(t)&=\underset{rr'ss'}{\sum}\,\overset{\infty}{\underset{n=0}{\sum}}\,\overset{n}{\underset{m=0}{\sum}}\left(\frac{D}{\omega_{c}}\right)^{n}\frac{\left(-1\right)^{m}}{\left(n-m\right)!m!}J_{rr'ss'}^{(\alpha)}e^{i\left(r-r'\right)\phi_{\alpha}}\nonumber\\
&\times\left\{e^{-iy^{\left(\alpha\right)}_{j,rs,m}\left(t-t_{0}\right)}\left[\overline{\Phi}\left(t-t_{0},\beta,-y_{j,r's',m}^{\left(\alpha\right)}\right)-\overline{\Phi}\left(t-t_{0},\beta,-\left(\overline{\varepsilon}_j-\mu\right)\right)\right]\right.\nonumber\\
&\left.+e^{i\Omega_{\alpha}\left(p_{1}\left(r-r'\right)+p_{2}\left(s-s'\right)\right)\left(t-t_{0}\right)}\Psi\left(\frac{1}{2}-\frac{\beta}{2\pi i}y^{\left(\alpha\right)}_{j,r's',m}\right)\right\},
\end{align}
%\textcolor{red}{Is it correct that only one $\Psi$ function appears in $K$? Looking at Eq. (27), seems the answer is yes. Then maybe remove one layer of parentheses around $\Psi$. -rt}
\begin{align}
P_{\alpha\alpha'jk}(t) & =\frac{e^{-i\left(\bar{\varepsilon}_{j}-\bar{\varepsilon}_{k}^{*}\right)\left(t-t_{0}\right)}}{\bar{\varepsilon}_{k}^{*}-\bar{\varepsilon}_{j}}\left[\Psi\left(\frac{1}{2}+\frac{\beta}{2\pi i}\left(\bar{\varepsilon}_{k}^{*}-\mu\right)\right)-\Psi\left(\frac{1}{2}-\frac{\beta}{2\pi i}\left(\bar{\varepsilon}_{j}-\mu\right)\right)\right]\nonumber\\
 & +\underset{rr'ss'}{\sum}\,\overset{\infty}{\underset{n=0}{\sum}\,}\overset{n}{\underset{m=0}{\sum}}\left(\frac{D}{\omega_{c}}\right)^{n}\frac{\left(-1\right)^{m}}{\left(n-m\right)!m!}J_{rr'ss'}^{(\alpha')}e^{i\left(r-r'\right)\phi_{\alpha'}}\mathcal{Z}_{\alpha\alpha'jk}^{rr'ss'nm}(t)\,,
\end{align}
and
\begin{align}
\mathcal{Z}_{\alpha\alpha'jk}^{rr'ss'nm}(t)=\xi_{jk,rr'ss'}^{\left(\alpha'\right)}(t)\left[\Psi\left(\frac{1}{2}-\frac{\beta}{2\pi i}y_{j,rs,m}^{(\alpha')}\right)-\Psi\left(\frac{1}{2}+\frac{\beta}{2\pi i}y_{k,r's',m}^{\left(\alpha\right)*}\right)\right]\nonumber\\
+\frac{1}{\bar{\varepsilon}_{k}^{*}-\bar{\varepsilon}_{j}}\left\{ e^{-iy_{j,rs,m}^{(\alpha')}\left(t-t_{0}\right)}\left[\bar{\Phi}\left(t-t_{0},\beta,-\overline{y}_{k,rs,m}^{(\alpha')*}\right)-\bar{\Phi}\left(t-t_{0},\beta,-y_{j,rs,m}^{(\alpha')}\right)\right]\right.\nonumber\\
+e^{-i\left(\bar{\varepsilon}_{j}-\bar{\varepsilon}_{k}^{*}\right)\left(t-t_{0}\right)}\left[\Psi\left(\frac{1}{2}-\frac{\beta}{2\pi i}\overline{y}_{k,rs,m}^{\left(\alpha'\right)*}\right)-\Psi\left(\frac{1}{2}-\frac{\beta}{2\pi i}y_{j,rs,m}^{(\alpha')}\right)\right]\nonumber\\
+e^{i\overline{y}_{k,r's',m}^{\left(\alpha\right)*}\left(t-t_{0}\right)}\left[\bar{\Phi}\left(t-t_{0},\beta,y_{k,r's',m}^{\left(\alpha\right)*}\right)-\bar{\Phi}\left(t-t_{0},\beta,\overline{y}_{j,r's',m}^{\left(\alpha\right)}\right)\right]\nonumber\\
\left.+e^{-i\left(\bar{\varepsilon}_{j}-\bar{\varepsilon}_{k}^{*}\right)\left(t-t_{0}\right)}\left[\Psi\left(\frac{1}{2}+\frac{\beta}{2\pi i}y_{k,r's',m}^{(\alpha')*}\right)-\Psi\left(\frac{1}{2}+\frac{\beta}{2\pi i}\overline{y}_{k,r's',m}^{(\alpha')}\right)\right]\right\} \nonumber\\
+\eta_{jk,rs,m}^{(\alpha')}\left\{ e^{-iy_{j,rs,m}^{(\alpha')}\left(t-t_{0}\right)}\left[\bar{\Phi}\left(t-t_{0},\beta,-\left(\bar{\varepsilon}_{j}-\mu\right)\right)-\bar{\Phi}\left(t-t_{0},\beta,-\overline{y}_{k,rs,m}^{\left(\alpha\right)*}\right)\right]\right.\nonumber\\
\left.+e^{-i\left(\bar{\varepsilon}_{j}-\bar{\varepsilon}_{k}^{*}\right)\left(t-t_{0}\right)}\left[\Psi\left(\frac{1}{2}-\frac{\beta}{2\pi i}\left(\bar{\varepsilon}_{j}-\mu\right)\right)-\Psi\left(\frac{1}{2}-\frac{\beta}{2\pi i}\overline{y}_{k,rs,m}^{\left(\alpha\right)*}\right)\right]\right\}\nonumber \\
+\eta_{kj,rs,m}^{\left(\alpha'\right)*}\left\{ e^{iy_{k,r's',m}^{\left(\alpha\right)*}\left(t-t_{0}\right)}\left[\bar{\Phi}\left(t-t_{0},\beta,\bar{\varepsilon}_{k}^{*}-\mu\right)-\bar{\Phi}\left(t-t_{0},\beta,\overline{y}_{j,r's',m}^{(\alpha')}\right)\right]\right.\nonumber\\
\left.+e^{-i\left(\bar{\varepsilon}_{j}-\bar{\varepsilon}_{k}^{*}\right)\left(t-t_{0}\right)}\left[\Psi\left(\frac{1}{2}+\frac{\beta}{2\pi i}\left(\bar{\varepsilon}_{k}^{*}-\mu\right)\right)-\Psi\left(\frac{1}{2}+\frac{\beta}{2\pi i}\overline{y}_{j,r's',m}^{(\alpha')}\right)\right]\right\} 
\end{align}
where we have introduced the following shortcuts alongside the ones in Eqs. \eqref{eq:J} and \eqref{eq:y}:
\[
\overline{y}_{j,rs,m}^{\left(\alpha\right)}=\bar{\varepsilon}_{j}-\mu-V_{\alpha}-\Omega_{\alpha}\left(p_{1}r+p_{2}s\right)+i\left(D+m\omega_{c}\right)\,,
\]
\[
\xi_{jk,rr'ss'}^{\left(\alpha'\right)}(t)=\frac{e^{-i\left(\bar{\varepsilon}_{j}-\bar{\varepsilon}_{k}^{*}\right)\left(t-t_{0}\right)}-e^{i\left(\Omega_{\alpha'}\left(p_{1}\left(r-r'\right)+p_{2}\left(s-s'\right)\right)\right)\left(t-t_{0}\right)}}{\bar{\varepsilon}_{k}^{*}-\bar{\varepsilon}_{j}-\Omega_{\alpha'}\left(p_{1}\left(r-r'\right)+p_{2}\left(s-s'\right)\right)}\,,
\]
\[
\eta_{jk,rs,m}^{\left(\alpha'\right)}=\frac{1}{\bar{\varepsilon}_{k}^{*}-\bar{\varepsilon}_{j}-V_{\alpha'}-\Omega_{\alpha'}\left(p_{1}r+p_{2}s\right)-i\left(D+m\omega_{c}\right)}\,,
\]

%\textcolor{red}{It seems $\overline{\eta}_{jk,rs,m}^{(\alpha')}=\eta_{kj,rs,m}^{(\alpha')*}$, so this property could be utilized above. For $y$ and $\overline{y}$ such symmetry does not hold. -rt}
Taking the limit in which the initial condition $t_0$ is set to the infinite past, and swapping the indices $r\leftrightarrow r'$, $s\leftrightarrow s'$, one obtains the expression in Eq. \eqref{eq:I_D_t}.
\end{widetext}

\bibliographystyle{apsrev4-1} 
\bibliography{Stochastic}

\end{document}